\documentclass[fleqn,usenatbib,useAMS]{mnras}

\usepackage{graphicx}	
\usepackage{amsmath}	
\usepackage{amssymb}	

\usepackage{xspace}
\usepackage{grffile}

\usepackage[dvipsnames]{xcolor}
\usepackage[normalem, normalbf]{ulem}

\newcommand{\dd}{\mathrm{d}}

\newcommand{\pow}[1]{\ifmmode{}^{#1}\else ${}^{#1}$\fi}

\newcommand{\cm}{\,\ifmmode{{\mathrm{cm}}}\else cm\fi}

\newcommand{\ergps}{\,{\rm erg}\,{\rm s}\ifmmode{}^{-1}\else${}^{-1}$\fi}
\newcommand{\Mpch}{\,{\rm Mpc}\,\ifmmode h^{-1}\else $h^{-1}$\fi}
\newcommand{\snru}{\,\ifmmode{\mathrm{Myr}^{-1}}\else Myr${}^{-1}$\fi}
\newcommand{\kms}{\,\ifmmode{\mathrm{km}\,\mathrm{s}^{-1}}\else km\,s${}^{-1}$\fi\xspace}

\newcommand{\cl}{\mathrm{cl}}
\newcommand{\lshatter}{\ifmmode{\ell_{\mathrm{shatter}}}\else $\ell_{\mathrm{shatter}}$\fi}
\DeclareMathOperator{\E}{\mathbb{E}}
\newcommand{\mytime}[1]{\ifmmode{{t_{\mathrm{#1}}}}\else $t_{\mathrm{#1}}$\fi}
\newcommand{\tcc}{\mytime{cc}}
\newcommand{\tcool}[1]{\mytime{cool,#1}}

\usepackage[T1]{fontenc}
\usepackage{ae,aecompl}

\usepackage{newtxtext,newtxmath}

\title[Evolution of turbulent, multiphase gas]{Survival and mass growth of cold gas in a turbulent, multiphase medium}

\author[Gronke, Oh, Ji, Norman]{
  Max Gronke${}^{1,2}$\thanks{E-mail: maxbg@jhu.edu, Hubble fellow},
  S. Peng Oh${}^{3}$,
  Suoqing Ji${}^{4,5}$,  Colin Norman${}^{1,6}$
\\
${}^{1}$ Department of Physics \& Astronomy, Johns Hopkins University, 
Bloomberg Center, 3400 N. Charles St., Baltimore, MD 21218, USA\\
${}^{2}$ Max Planck Institut fur Astrophysik, Karl-Schwarzschild-Straße 1, D-85748 Garching bei München, Germany\\
    ${}^{3}$ Department of Physics, University of California, Santa Barbara, CA 93106, USA\\
    ${}^{4}$ Shanghai Astronomical Observatory, Chinese Academy of Sciences, Shanghai  200030, China \\
    ${}^{5}$ TAPIR \& Walter Burke Institute for Theoretical Physics, Caltech, Pasadena CA 91125, USA\\
  ${}^{6}$   Space Telescope Science Institute, Baltimore, MD 21218, USA
}

\date{Draft from \today}

\pubyear{2021}

\begin{document}
\label{firstpage}
\pagerange{\pageref{firstpage}--\pageref{lastpage}}
\maketitle

\begin{abstract}
  Astrophysical gases are commonly multiphase and highly turbulent. In this work, we investigate the survival and growth of cold gas in such a turbulent, multi-phase medium using three-dimensional hydrodynamical simulations.
  Similar to previous work simulating coherent flow (winds), we find that cold gas survives if the cooling time of the mixed gas
  is shorter than the Kelvin-Helmholtz time of the cold gas clump (with some weak additional Mach number dependence). However, there are important differences. 
  Near the survival threshold, the long-term evolution is highly stochastic, and subject to the existence of sufficiently large clumps. In a turbulent flow, the cold gas continuously fragments, enhancing its surface area. This leads to exponential mass growth, with a growth time given by the geometric mean of the cooling and the mixing time. The fragmentation process leads to a large number of small droplets which follow a scale-free $\mathrm{d} N/\mathrm{d} m \propto m^{-2}$ mass distribution, and dominate the area covering fraction. Thus, whilst survival depends on the presence of large `clouds', these in turn produce a `fog' of smaller droplets tightly coupled to the hot phase which are probed by absorption line spectroscopy. We show with the aid of Monte-Carlo simulations that the simulated mass distribution emerges naturally due to the proportional mass growth and the coagulation of droplets. We discuss the implications of our results for convergence criteria of larger scale simulations and observations of the circumgalactic medium.
 \end{abstract}


\section{Introduction}
\label{sec:intro}
Turbulent, multiphase gases are extremely common in astrophysics. We find them in the interstellar-, circumgalactic-, intracluster-, and even intergalactic medium (ISM, CGM, ICM and IGM, respectively; for reviews, see, e.g., \citealp{Klessen2014,Kravtsov2012a,Tumlinson2017,Meiksin2009}).
Understanding the dynamics of such multi-phase gases is crucial.

It is essential to understand the phase structure of such gases and the mass, momentum and energy exchange between phases, which can be mediated by physical mechanisms such as radiative cooling, heating, and thermal conduction. In a static medium, this exchange is relatively well understood \citep{Field1965,Waters2018,Das2020}. However, in a dynamic situation, hydrodynamical instabilities and turbulence can mix the phases, and thus can quickly dominate the mass, momentum and energy transfer rates between hot and cold medium. As astrophysical systems are rarely static -- and in fact often highly turbulent -- this regime is of great interest.

Several studies have been carried out with the broad goal of characterizing the mass transfer rate between different phases, for instance, focusing on thermal instabilities \citep{Sharma2010,Mccourt2012,Sharma2012,Voit2015}, turbulent mixing layers \citep{Begelman1990,Ji2018,Fielding2020,Tan2020}, the interaction of a hot wind with a cold cloud \citep[e.g.,][]{Klein1994,Mellema2002,Scannapieco2015a,Bruggen2016,Schneider2016,Armillotta2016,Gronke2018,Sparre2018,Li2019a,Gronke2020,Kanjilal2020,Abruzzo2021,Farber2021}, or more complex multiphase geometries \citep{2015ApJ...802...99K,2021MNRAS.504.1039R,2020MNRAS.499.2173B,2021MNRAS.506.5658B}.

In this paper, we focus on the dynamics of a cold cloud embedded in a turbulent hot medium, as can be found in essentially all the astrophysical systems mentioned above. We focus on the regime where the cooling time of the hot medium is long. We do not study thermal instability in a turbulent medium, which can also lead to multi-phase structure and has been the focus of several previous studies \citep{Hennebelle1999,Gazol2001,Kritsuk2002,Saury2014,Kobayashi2020}.

The paper is structured as follows. We will first set the stage in Sec.~\ref{sec:analytics} where we discuss the analytic expectations. Then, we will describe in Sec.~\ref{sec:methods} our numerical setup before we present its results in Sec.~\ref{sec:results}. We discuss our findings in Sec.~\ref{sec:discussion} before we conclude in Sec.~\ref{sec:conclusion}. Videos visualizing our results can be obtained at \url{http://max.lyman-alpha.com/multiphase-turbulence}.

\begin{figure*}
  \centering
  \includegraphics[width=\linewidth]{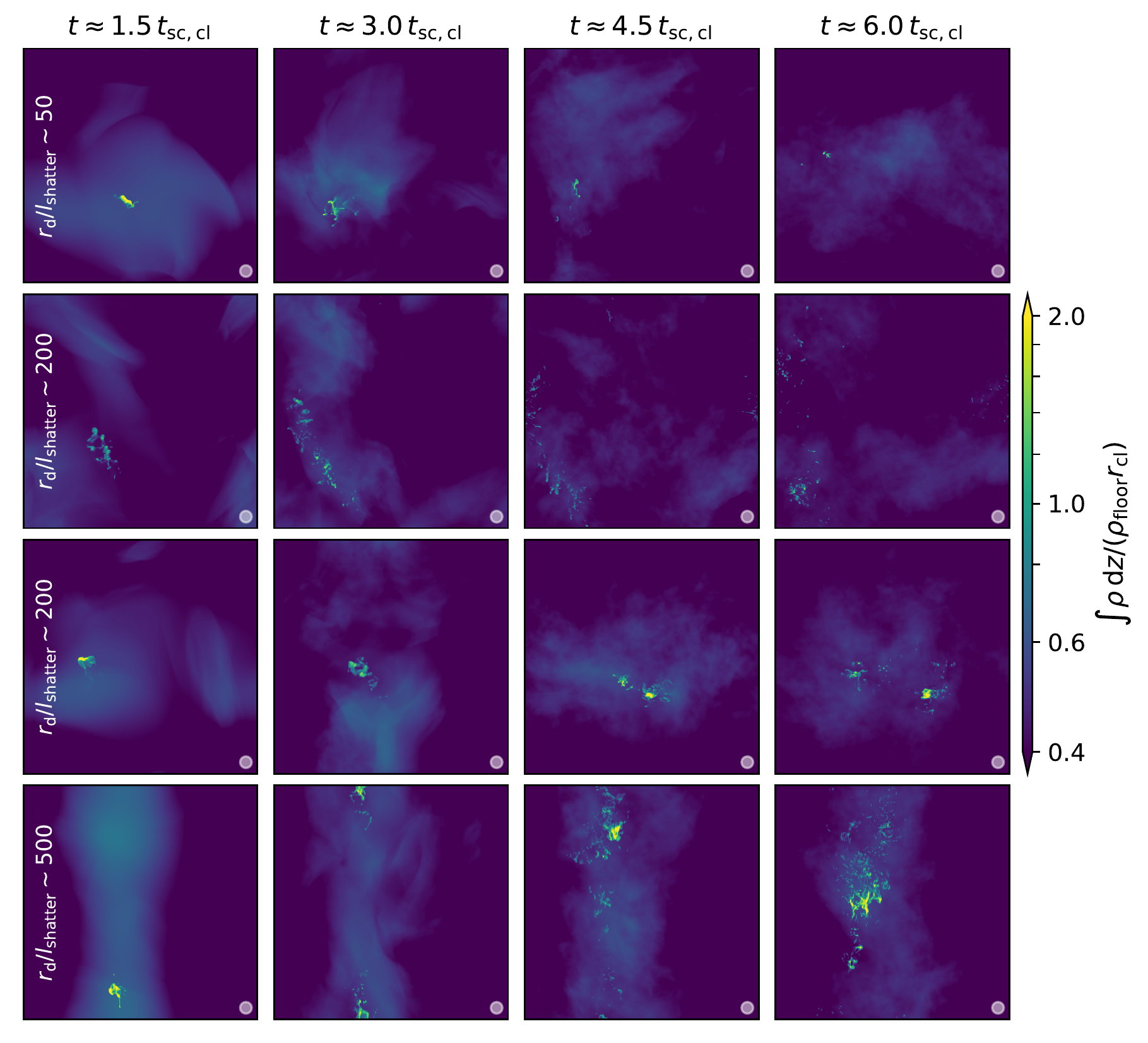}
  \caption{Density projections of individual droplets of different sizes (from $r_{\rm d}\sim 50\lshatter$ to $r_{\rm d}\sim 500\lshatter$ from top to bottom row) and overdensity $\chi\sim 100$ in a turbulent medium with $\mathcal{M}\sim 1$. The small droplet is dispersed quickly whereas the larger one manages to survive. The white circle indicated the initial droplet size. Videos showing these simulations can be seen at \url{http://max.lyman-alpha.com/multiphase-turbulence}.}
  \label{fig:multi2d_3dturb_N1_rdvar}
\end{figure*}

\section{Analytic considerations}
\label{sec:analytics}

In this section, we review characteristic length scales, and discuss expectations for cold gas growth rates. 

We consider a cloud of size $r_\cl$, overdensity $\chi$ and temperature $T_{\rm cold}$ embedded in a hot medium with temperature $T_{\rm hot}$ which has an rms turbulent velocity $v_{\rm turb}=\mathcal{M}c_{\rm s,hot}$.
The `cloud crushing problem' where the cloud is subject to a uniform wind has been heavily studied. There, hydrodynamical instabilities will destroy the cold gas on a timescale of the order of the Kelvin-Helmholtz, Rayleigh-Taylor, or shock-crossing time of the cloud $t_{\rm cc}\sim \chi^{1/2}r_\cl/v_{\rm turb}$ \citep{Klein1994}. 
From our previous work \citep{Gronke2018}, we expect the cloud to survive if
\begin{equation}
  \label{eq:tcoolmix_crit}
  t_{\rm cool,mix}< \alpha t_{\rm cc},
\end{equation}
that is, when the cooling time of the mixed gas $t_{\rm cool}(T_{\rm mix}, n_{\rm mix})$ with $T_{\rm mix}\sim \sqrt{T_{\rm hot}T_{\rm cold}}$,  $n_{\rm mix}\sim \sqrt{n_{\rm hot}n_{\rm cold}}$ is smaller than the cloud crushing time. In this `wind-tunnel' setup, we found Eq.~\eqref{eq:tcoolmix_crit} to hold with $\alpha\sim 1$ (\citealp{Gronke2018}; see also discussion of this criterion in \citealp{Kanjilal2020} and references therein) but as we will see the dynamics is more complex in a turbulent multiphase medium and thus we leave $\alpha$ as a free fudge parameter for now.
Eq.~\eqref{eq:tcoolmix_crit} can be rewritten as a geometrical criterion which states that clouds larger than a characteristic size survive the ram pressure acceleration process \citep{Gronke2018,Gronke2019}: 
\begin{equation}
R > r_{\rm crit,w} \sim \frac{v_{\rm wind}\tcool{mix}}{\chi^{1/2}} \alpha^{-1} \approx 2 \, {\rm pc} \ \frac{T_{\rm cl,4}^{5/2} \mathcal{M}_{\mathrm{wind}}}{P_{3} \Lambda_{\rm mix,-21.4}} \frac{\chi}{100}\alpha^{-1}
\end{equation}
where $T_{\rm cl,4} \equiv (T_{\rm cl}/10^{4} \, {\rm K})$, $P_{3} \equiv n T/(10^{3} \, {\rm cm^{-3} \, K})$, $\Lambda_{\rm mix,-21.4} \equiv \Lambda(T_{\rm mix})/(10^{-21.4} \, {\rm erg \, cm^{3} \, s^{-1}})$, $\mathcal{M}_{\mathrm{wind}}$ is the Mach number of the wind, and we write $v_{\mathrm{wind}} = c_{\mathrm{s,wind}} \mathcal{M}_{\rm wind} \sim c_{\mathrm{s,cl}}\mathcal{M}_{\rm wind}\chi^{1/2}$. This implies that for typical galactic conditions clouds larger than parsec size usually fulfill the requirement. The scale $r_{\rm crit,w}$ can be compared to the characteristic length scale of cooling induced fragmentation $\lshatter \sim c_{\mathrm{s,cl}}t_{\mathrm{cool,cl}}$ \citep{McCourt2016,Gronke2020} to give
\begin{equation}
\frac{r_{\rm crit,w}}{\lshatter} \approx 50 \ \mathcal{M}_{\rm wind} \frac{\chi}{100} \left(
\frac{\Lambda(T_{\rm cl})/\Lambda(T_{\rm mix})}{0.5} \right) \alpha^{-1}.
\label{eq:rcrit_lshatter} 
\end{equation}
where quantities are evaluated at values appropriate for our fiducial simulation. Thus, a cloud should be significantly larger than the characteristic fragmentation scale to exhibit this behavior.
 Note that we use $r_{\rm crit,w}$ for the `wind tunnel' solution to differ from the survival scale $r_{\rm crit}$ in a turbulent medium investigated in this work.

If a cloud survives, we expect the cold gas to grow in mass due to continuous cooling which we characterize by
\begin{equation}
  \label{eq:mdot}
  \dot m \sim v_{\mathrm{mix}} A_\cl \rho_{\mathrm{hot}}
\end{equation}
with an effective cold gas surface area $A_\cl$ (see below), and a surrounding hot gas density $\rho_{\mathrm{hot}}$. The dynamics of the mixing layer is dominated by turbulence.
The mixing velocity follows (in the `fast cooling regime', when the cooling time is shorter than the eddy turnover time) the scaling \citep{Gronke2019,Fielding2020,Tan2020}
\begin{equation}
  \label{eq:vmix_prop}
  v_{\rm mix}\propto (u')^{3/4} \left( \frac{L_{\rm cold}}{t_{\rm cool,c}} \right)^{1/4}
\end{equation}
where $u'$ is the turbulent velocity in the cold medium and depends on the specific setup\footnote{For a plane-parallel shearing layer \citep{Tan2020} found $u'\propto \mathcal{M}^{4/5}$.}. Here, $t_{\rm cool}$ and the integral scale of turbulence $L_{\rm cold}$ are also evaluated for the cold medium.
In summary, generally the mass transfer between the phases depends on the cold gas surface area and the mixing rate per unit area.

The former dependence is intuitive. In combustion, the reaction rate is proportional to the reactant surface area; likewise here, mass growth (where hot `fuel' is converted via a cold `reactant' to cold `ashes') is proportional to the cold gas surface area. In our wind-tunnel simulations, the `effective' surface area $A_\cl$ of a monolithic cloud in equation \ref{eq:mdot} followed a simple scaling $A_\cl \sim (m/\rho_{\cl})^{2/3}$.  However, in turbulent setup where cold gas is constantly fragmenting to smaller scales, the surface area can grow faster with mass. The cold-hot gas interface has a fractal geometry \citep{Fielding2020}, and it is well-known that fractals have surface areas which grow faster than the Euclidean expectation of $A \propto V^{2/3}$, as inferred from e.g., fractal respiratory organs. For instance, one possible ansatz is that $A \propto V^{D/3}$, where $2 < D < 3$ is the fractal dimension \citep{Barenblatt1983}. Another limiting case is when large clouds continuously fragment down to some scale $\tilde{r}$. If so, the surface area is simply proportional to the number of such small clouds, so that $A \propto m$. These differences are important: for $A_\cl \propto m^{2/3}$ one obtains a power law solution of the form $m(t)=(1+a t)^3$ (with $a$ being a combination of the model parameters) whereas for $A_\cl\propto m$ it follows $m \sim m_0 \exp(t/t_{\rm grow})$ with $t_{\rm grow}\sim \chi \tilde r / v_{\rm mix}$ where $\tilde r$ is the fragmentation size. We shall later see that in a turbulent scenario the mass growth rate is indeed exponential.

How the turbulent surface area $A_{\rm T}$ scales with physical parameters is a thorny problem which is a central focus of the turbulent combustion literature. Many different scalings have been proposed, each with experimental support in different regimes (see discussion in \citealp{Tan2020} and references therein). \citet{Tan2020} side-stepped this problem by arguing that a turbulent medium should have a net cooling time given by the geometric mean of the eddy turnover and cooling time of the cold medium (cf. equation~(25) of \citealp{Tan2020}): 
\begin{align}
  \label{eq:tcool_eff}
  \tilde t_{\rm cool}\sim \left( \frac{l}{u'} t_{\rm cool,c}\right)^{1/2}
\end{align}
where $l$ and $u'$ are the characteristic length scale and turbulent velocity in the \textit{cold medium}, respectively. This can be thought of as the geometric mean of the elastic and inelastic timescales for a fluid element, similar to random walk timescales for a photon in the presence of both absorption and scattering, which gives rise to an effective optical depth $\tau_{\rm eff} \sim \sqrt{\tau_{\rm abs} \tau_{\rm scatter}}$ or an electron in the presence of both Coulomb (elastic) and atomic (inelastic) scattering, which gives rise to the Field length $\lambda_{\rm F} \sim \sqrt{\lambda_e \lambda_{\rm cool}}$. This ansatz was supported by detailed simulations of turbulent mixing layers. This implies a mass growth time: 
\begin{equation}
  \label{eq:tgrow_TML_full}
  t_{\rm grow}\equiv \frac{m}{\dot m} \sim \chi \tilde t_{\rm cool} \sim \chi\mathcal{M}^{-1/2} \left( \frac{l}{\lshatter} \right)^{1/2} \left(\frac{l}{L_{\rm box}} \right)^{-1/6} t_{\rm cool,c}.
\end{equation}
We have assumed that -- as seen in hydrodynamic simulations -- the turbulent pressure is continuous across the interface $\rho_c (u^{\prime})^2 \approx \rho_h v_{\rm turb}^{2}$. Thus, the cold gas turbulent velocity is $u'\sim v_{\rm turb,hot}(l)/\chi^{1/2}=\mathcal{M}c_{\rm s,c} (l/L_{\rm box})^{1/3}$ at the scale $l$. The scale $l$ is the length scale characterizing the cold-hot gas interface. Initially, this is given by the cloud-size; later on, a combination of mass growth, fragmentation and coagulation renders this length scale more ambiguous. In principle, at late times, $l$ approaches the driving scale of turbulence $L_{\rm box}$. This is true of mixing layer simulations, where cold gas fills the box.  
Fortunately,
Eq.~\eqref{eq:tgrow_TML_full} implies that $t_{\rm grow}$ only has a weak $t_{\rm grow}\propto l^{1/3}$ dependence on the lengthscales $l$. We have found that with our limited simulation domain and the boundary conditions we use, $l\sim r_\cl$ is a good approximation throughout. Furthermore, since $t_{\rm grow} = m/\dot{m} \propto l^{1/3} \propto m^{1/9}$ is roughly constant\footnote{The cooling time of gas in the cold medium $t_{\rm cool,c}$ depends only on pressure and metallicity, which we assume to be roughly constant.} (and, equivalently $v_{\rm mix}$), we expect exponential mass growth $m \propto {\rm exp}(t/t_{\rm grow})$. \\

We will explore the validity of both the survival estimate and the simple models for mass growth using numerical simulations.

\begin{figure}
  \centering
  \includegraphics[width=\linewidth]{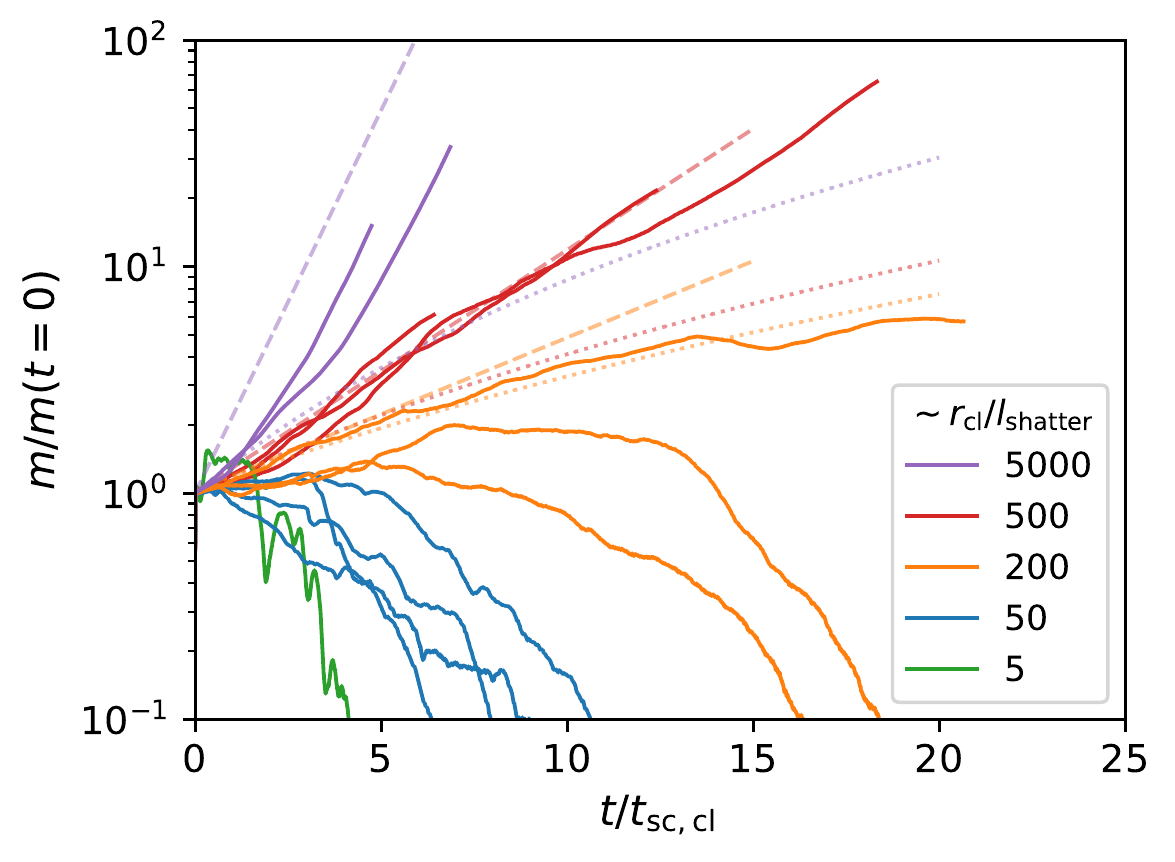}
  \caption{Mass evolution of simulations with different initial cloud sizes. In all simulations, we placed $N_{\rm d}=1$ cloud in a turbulent $\mathcal{M}\sim 1$ medium. The dotted (dashed) lines in the corresponding colors show the solutions of our analytic estimates for monolithic (fragmented) growth. Note that for these parameters $r_{\rm crit,w}/l_{\rm shatter}\sim 70$.}
  \label{fig:mevo_rdvar}
\end{figure}

\begin{figure}
  \centering
  \includegraphics[width=\linewidth]{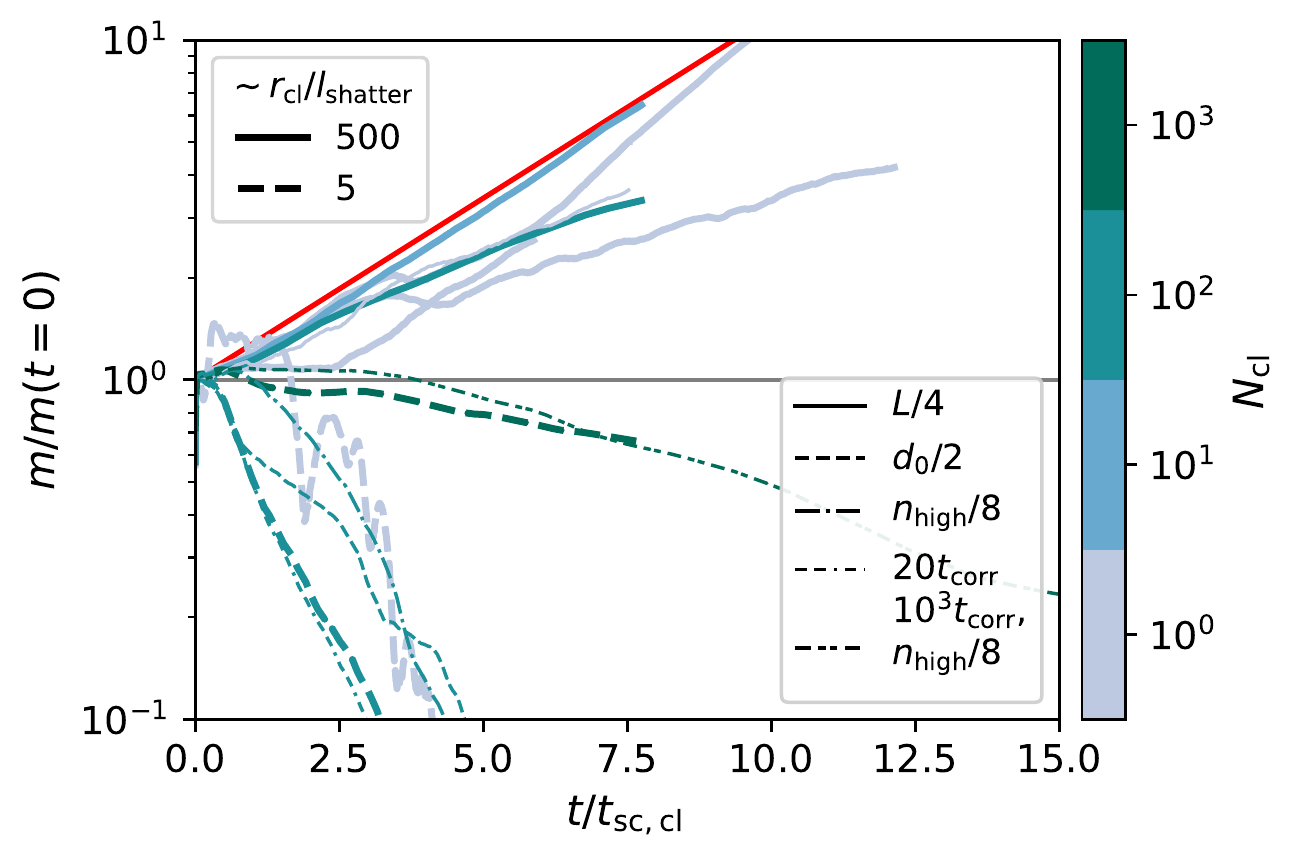}
  \caption{Mass evolution of an ensemble of droplets in a turbulent medium (with $\mathcal{M}\sim 1$). The fiducial values of the runs were $\chi\sim 100$ and the droplets were distributed in a sphere of radius $d_0\sim 15 r_{\rm d}$ in a periodic box with side length $L_{\rm Box}=20 r_{\rm d}$ and $512^3$ cells. The thinner lines show simulations where we changed these fiducial values slightly. The red solid line shows our analytic estimate for fragmented growth (i.e. equation \ref{eq:tgrow_TML_full}) with a fudge factor of $0.5$. Note that the smaller droplets are destroyed whereas the larger droplets (solid lines) survive, and in the latter case the mass growth rate does not depend on the initial number of droplets.}
  \label{fig:coag3d_turb}
\end{figure}

\section{Numerical Methods}
\label{sec:methods}

For our hydrodynamical simulation, we use the \texttt{Athena++} code \citep{Stone2020}.
We use the HLLC Riemann solver, second-order reconstruction with slope limiters in the primitive variables, and the van Leer unsplit integrator \citep{Gardiner2008}. We implemented the \citet{Townsend2009} cooling algorithm which allows for fast and accurate computations of the radiative losses. We adopt a solar metallicity cooling curve to which we fitted broken a power-law, and a temperature floor $T_{\mathrm{floor}}=4\times 10^4\,$K\footnote{We use this value to be comparable with earlier studies \citep{2015MNRAS.449....2M,Gronke2018} and to ensure the gas is fully ionized since we assume collisional ionization equilibrium cooling rates.}. Note that $\lshatter$ is evaluated at this temperature floor. 

For this work, we place $N_\cl$ droplets with approximate size $r_{\mathrm{d}}$\footnote{To avoid the carbuncle instability, the clouds are not perfectly spherical but rather potato shaped. We furthermore introduce density perturbations on the $\sim 1\%$ level on the entire domain.}, overdensity $\chi_{\mathrm{i}}$, temperature $T_{\mathrm{cl,init}}$ and the same pressure as the background randomly within a radius $d_{0}$ which we continuously stir\footnote{For visualizations of the initial conditions, we refer the reader to the animations available at \url{http://max.lyman-alpha.com/multiphase-turbulence}.}. Specifically, we impose an initial, turbulence field in the box (for both the hot medium as well as cold clouds), and then stir continuously with the wavenumbers $n_{\rm low}$ to $n_{\rm high}$ (i.e., we stir at scales $k=2 \pi n / L_{\rm box}$ with $n_{\rm low}<n < n_{\rm high}$) with a initial Kolmogorov spectrum and a correlation time $t_{\rm corr}$ and a ratio of solenoidal to compressive  components $f_{\mathrm{shear}}$ to yield an approximately constant turbulent Mach number $\mathcal{M}\equiv v_{\mathrm{turb}}/c_{\rm s, hot}$. Our fiducial values are $n_{\rm high} = 2$, $n_{\rm low} = 0$ (i.e., we stir at the largest scale), $d_{0}=15 r_{\mathrm{d}}$, $T_{\rm cl,init} = 8\times 10^4\,$K, $\chi_{\rm init}=50$, $f_{\mathrm{shear}}\sim 1/3$, $t_{\rm corr}\sim 2 t_{\rm eddy}\sim 2 L / v_{\rm turb}$ where $L$ is the boxsize for which we usually choose $40 r_{\mathrm{cl}}$ (yielding an initial cold gas mass fraction of $m_{\rm cold} / m_{\rm total}\sim 0.003$ for $N_\cl = 1$, $\chi\sim 100$, and $t_{\rm eddy}\sim L / (\mathcal{M} \chi^{1/2}c_{\rm s,cold})\sim 4/\mathcal{M} t_{\rm sc,cl}$) where $t_{\rm sc,cl}$ is the sound crossing time of a cloud at its floor temperature.
See Appendix~\ref{sec:app_turb_params} for the effect of the driving parameters on the outcome. 
As the cooling time of the cold gas is short compared to all other timescales,
the clouds quickly settle at $T_\cl = T_{\rm floor}$. We found this initial perturbation to have no effect on the evolution (for sufficiently stirred boxes), see Appendix~\ref{sec:app_turb_params} for details.

For all our setups, we strive to resolve the cold gas by at least $\sim 16$ cells to ensure a convergent behavior \citep[see][for an extensive discussion on resolution requirements]{Tan2020}. We do, however, increase the resolution to $\sim 64$ cells and decrease it to $\sim 4$ cells to check this explicitly in some cases (see \S~\ref{sec:results_stochastic} and discussion in \S~\ref{sec:disc_scale} for details on convergence). Furthermore, we employ periodic boundary conditions.

A small subset of our simulations -- the ones using Lagrangian tracer particles (\S~\ref{sec:droplet_dists}) -- were carried out using the \texttt{FLASH} code \citep{fryxell00}. The \texttt{FLASH} code \citep{fryxell00} solves
 the equations of inviscid hydrodynamics with a directionally unsplit hydro solver, based on a finite-volume, high-order Godunov method  \citep{tzeferacosetal12, lee2013solution}. Also here, cooling is included as a source term with the ``exact'' cooling algorithm described in \citet{Townsend2009}, which preserves the accuracy of our cooling calculations. We use the same parameters as described above with the difference that we first apply the driving to the simulation box until turbulence becomes fully developed (for $\sim 2.4\,t_{\rm eddy}$) and statistically stable state is reached, with a spatially-averaged Mach number of $\mathcal{M}\sim 0.4$. We then manually add a droplet by setting the density and temperature in a spherical region of size $r_\cl$ to be $\chi_{\rm init}$ times higher and lower, respectively.

\begin{figure}
  \centering
  \includegraphics[width=\linewidth]{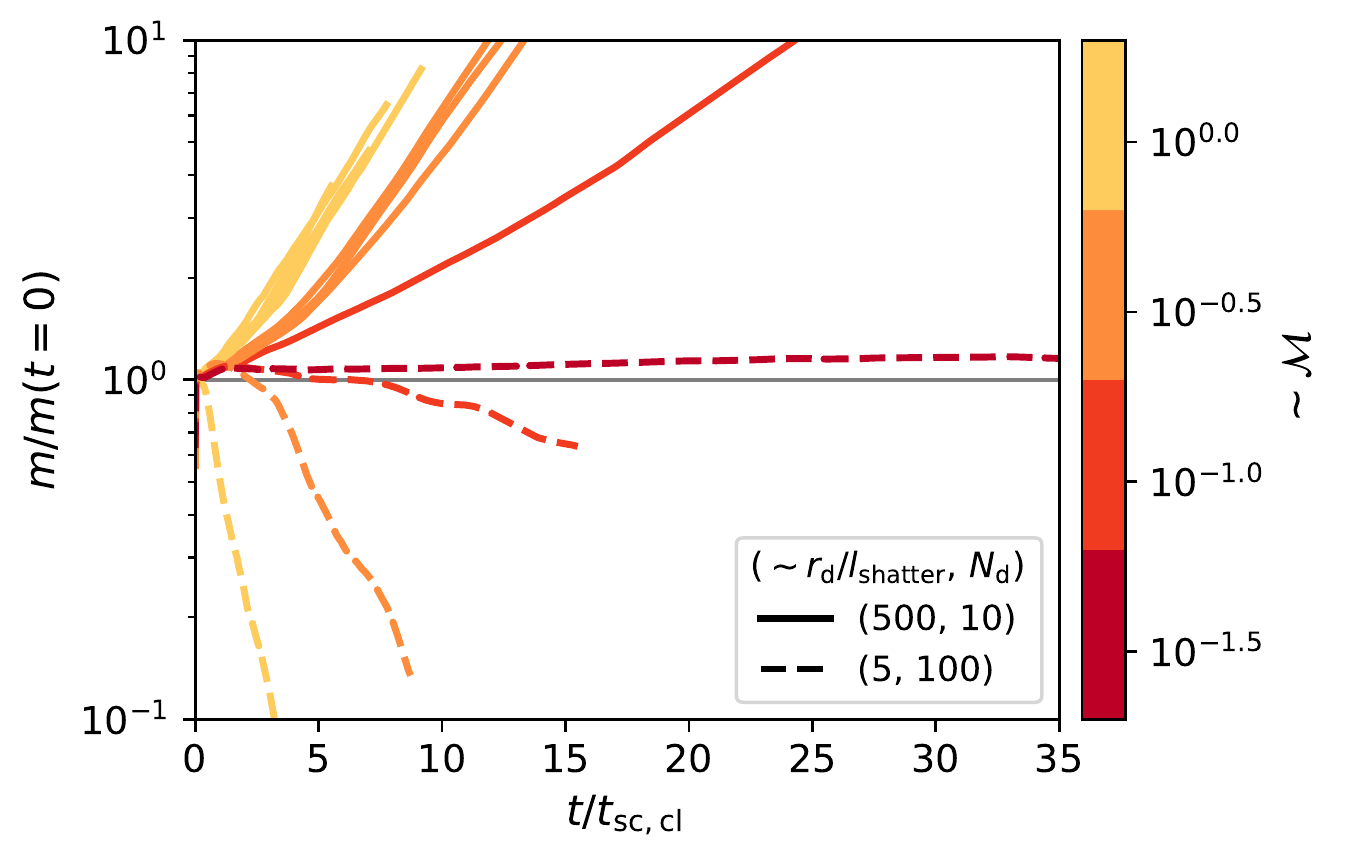}
  \caption{Mass evolution of an ensemble of droplets with $\chi_{\rm floor}\sim 100$ in a turbulent medium with varying energy injection. An increase of turbulent energy leads to a increase in mass growth and destruction rates for the large and small droplets, respectively.}
  \label{fig:coag3d_turb_Evary}
\end{figure}

\begin{figure}
  \centering
  \includegraphics[width=\linewidth]{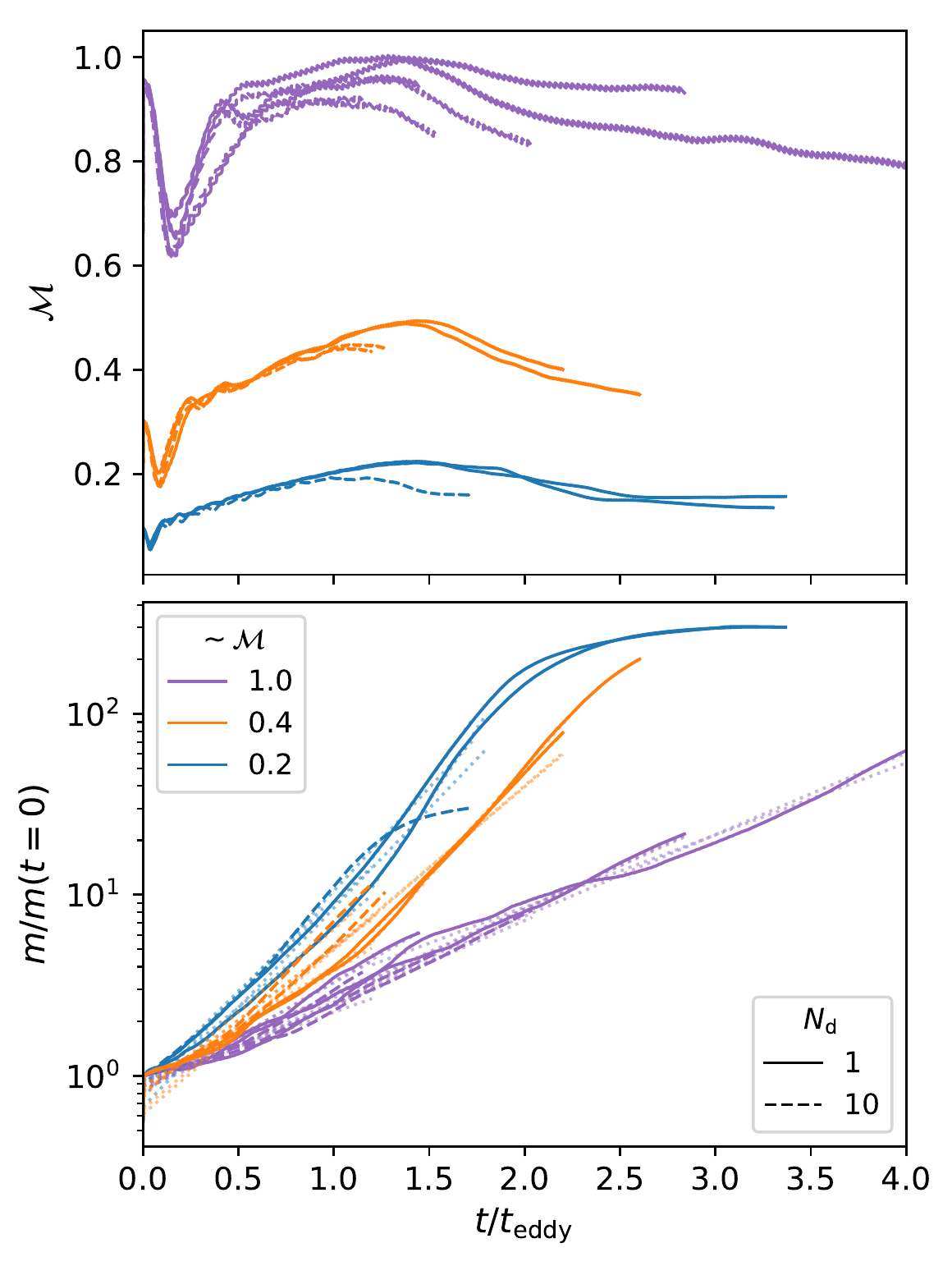}
  \caption{Mach number and mass evolution of (an ensemble) of droplets with $r_{\mathrm{d}}\sim 500\lshatter$ and $\chi\sim 100$ in a turbulent medium with varying energy injection. The shaded dotted lines represent the exponential fits detailed in Fig.~\ref{fig:coag3d_turb_Machvar_expfits}. Note that the time was normalized here to $t_{\rm eddy}\sim L_{\rm box} / (c_{\rm s,hot}\mathcal{M})$. As before, curves with the same color and linestyle show runs with different random seeds. }
  \label{fig:coag3d_turb_Evary_rseed}
\end{figure}

\section{Results}
\label{sec:results}

\subsection{Impact of the cloud size on cold gas survival and growth}
\label{sec:results_cloudsize}

Figure~\ref{fig:multi2d_3dturb_N1_rdvar} shows density projections for simulations of different initial cloud sizes with a fixed turbulent velocity of $\mathcal{M}\sim 1$, and overdensity of $\chi\sim 100$. While the smallest cloud ($r_\cl \sim 50\lshatter$; top row of Fig.~\ref{fig:multi2d_3dturb_N1_rdvar}) is destroyed rapidly, the bigger cloud ($r_\cl \sim 500\lshatter$) survives. The central two rows of Fig.~\ref{fig:multi2d_3dturb_N1_rdvar} show both simulations with $r_\cl\sim 200\lshatter$ but different random seeds (i.e., a different stirring pattern). In one, the cold gas survives, in the other it dies.

Figure~\ref{fig:multi2d_3dturb_N1_rdvar} also illustrates that the surviving cloud (lowest row) does not stay compact. Instead it is fragmented into many small droplets which will eventually fill the entire simulation domain. We will study the droplet census and its impact on the mass growth in \S~\ref{sec:detail_survival},

Figure~\ref{fig:mevo_rdvar} shows more quantitatively the cold gas mass evolution of these (and additional) simulations. The same picture emerges: all $r_{\mathrm{d}} / \lshatter \sim 500$ clouds grow in mass and clouds $r_{\mathrm{d}}\lesssim 50\lshatter$ are destroyed on a timescale of a few $t_{\rm cc}$ (which is for the simulations shown in Fig.~\ref{fig:mevo_rdvar} $\sim t_{\rm sc,cl}$ as there $\mathcal{M}\sim 1$). Note that Fig.~\ref{fig:mevo_rdvar} shows several simulations with the same runtime parameters but different random seeds which can lead to a very different outcome of the simulation. We discuss this stochasticity effect further in  \S~\ref{sec:results_stochastic}.

Figure~\ref{fig:mevo_rdvar} also shows the mass growth rate as predicted in \S~\ref{sec:analytics}. Specifically, we show both the evolution expected from fragmented (exponential) and monolithic growth (power-law) with  dashed and dotted lines, respectively; note we used a fudge factor of $\sim 0.5$ for the fragmented growth for $t_{\rm grow}$. The fragmented growth (leading to $m \propto \exp(a t)$) seems to fit the runs with $r_\cl/\lshatter \gtrsim 500$ better (here, as below, we use $\tilde r\sim r_\cl$) whereas the `monolithic growth' captures the evolution of the surviving $r_\cl \sim 200 \lshatter$ run better. Interestingly, this seems to agree with visual inspection in which the clump in this run is growing a ``tail'' (akin to the cloud-crushing simulations mentioned above), whereas the run with the larger cloud is more fragmented. We will investigate this further in \S~\ref{sec:detail_survival}.

\subsection{(Non-)shielding of multiple clouds}
\label{sec:results_Nclouds}

We saw in the last section that clouds with sizes of $5\lshatter$ and $500\lshatter$ placed in a $\mathcal{M}\sim 1$ box die and survive, respectively. What happens if we place many small $\sim 5\lshatter$ clouds in the box forming an effective cloud of a larger size? Will they then survive due to the shielding of each other from ram pressure?
In Fig.~\ref{fig:coag3d_turb} we show the mass growth rate of an ensemble of droplets (with the number color coded) with size $r_{\rm d}\sim 500 \lshatter$ and $r_{\rm d}\sim 5\lshatter$ in solid and non-solid colored lines, respectively. As before, one can note that the smaller droplets get destroyed whereas the large droplets grow, even when placing $N_{\mathrm{d}}=1000$ small droplets in the box. We also explored different driving parameters  (shown with different linestyles in Fig.~\ref{fig:coag3d_turb}, also see Appendix \ref{sec:app_turb_params}) -- but the small droplets still disintegrate. In a turbulent setup with continual direct forcing throughout the box, the `shielding' of cloudlets does not seem to prolong their survival times significantly. However, this may be different in other flow geometries. For instance, \citet{McCourt2016} and \citet{Forbes2019} study this shielding effect in a laminar flow \citep[see also][]{2020MNRAS.499.2173B,2021MNRAS.506.5658B} and find that although the cold gas mass is decreasing\footnote{Part of this is due to cold gas leaving the simulation domain, so the overall effect is unclear. Also, these simulations were run in the limit of large cold gas mass fraction, which enhances the effects of shielding.}, its lifetime can be prolonged.
This suggests that shielding effects are only important in bulk flows.
Studying this effect in a complex flow may be an interesting avenue for future work.

On the other hand, for simulations with cold gas mass growth, the normalized mass growth does not depend on $N_{\rm d}$. The solution of  Eq.~\eqref{eq:tgrow_TML_full}
is shown as red line in Fig.~\ref{fig:coag3d_turb} (using a fudge factor of $0.5$ for $t_{\rm grow}$).

\subsection{Variation of the turbulent driving energy}
\label{sec:results_Eturb}

\begin{figure}
  \centering
  \includegraphics[width=\linewidth]{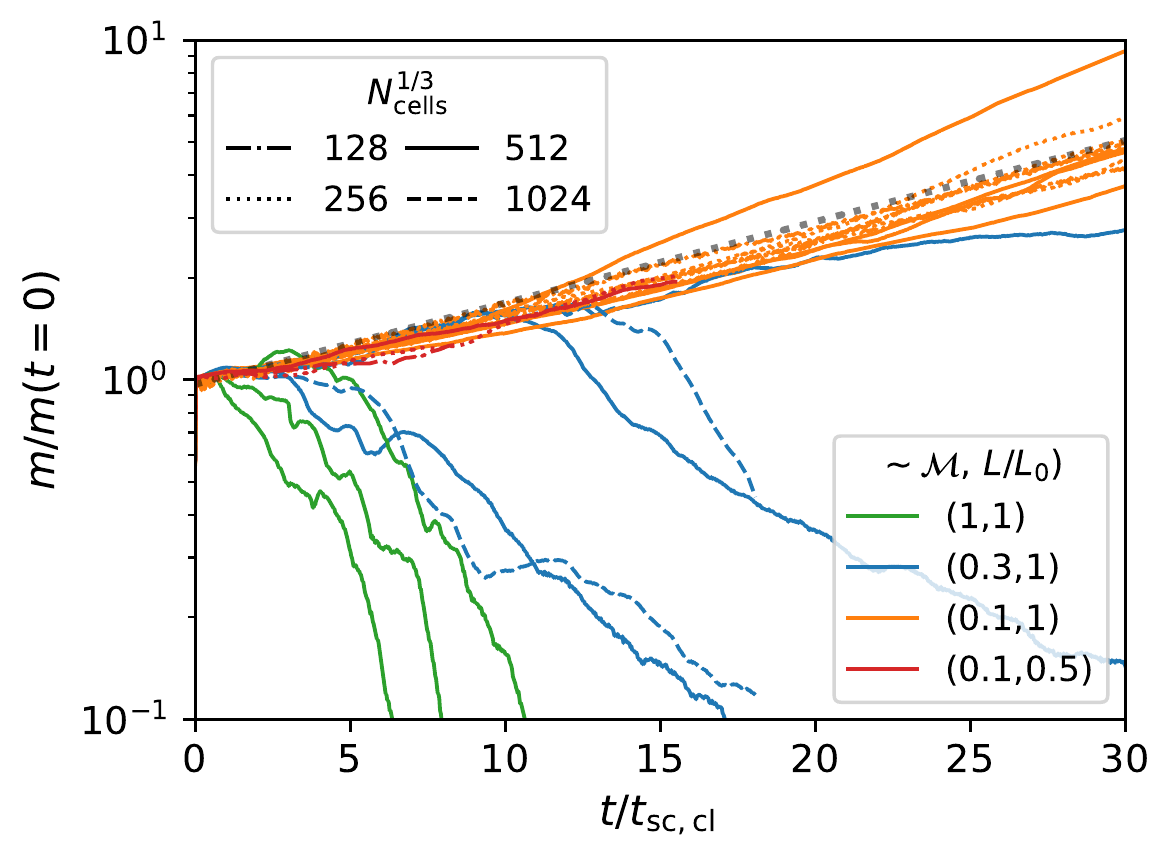}
  \caption{Mass evolution of a single droplet of size $r_{\rm d}\sim 50\lshatter$ ($\chi\sim 100$) in a turbulent medium with varying energy injection. Note that for the $\mathcal{M}\sim 1$ run (green lines)  $t_{\rm cool,mix}/\tcc \sim 1$ whereas for the $\mathcal{M}\sim 0.1$ runs (orange and red lines) $t_{\rm cool,mix}/\tcc \sim 0.2$. The latter gain mass independent of resolution. The $\mathcal{M}\sim 0.3$ runs (blue lines) are at the boundary of survival and the outcome is stochastic. The dotted gray line shows the model with a fudge factor of $0.5$ in $t_{\rm grow}$. }
  \label{fig:coag3d_turb_rd50}
\end{figure}

\begin{figure}
  \centering
  \includegraphics[width=\linewidth]{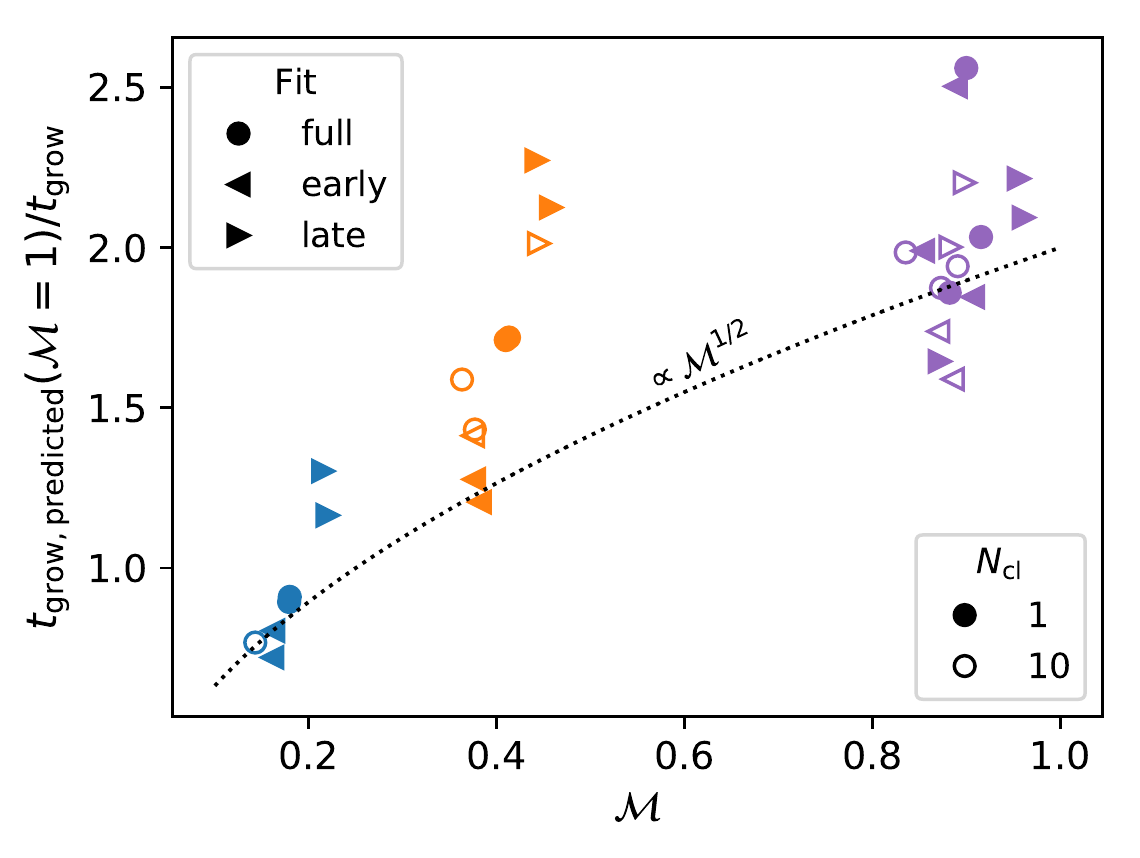}
  \caption{Comparison of the exponential mass growth rates from the simulations shown in Fig.~\ref{fig:coag3d_turb_Evary_rseed}. All the fits were carried out over the time range for which the cold gas mass fraction in the box is $<30\%$. In addition, the `early' and `late' fits are for $t< 1.2 t_{\rm eddy}$ and $t>1.2 t_{\rm eddy}$, respectively. The scaling with Mach number follows Eq.~\eqref{eq:tgrow_TML_full} with a fudge factor of $0.8$.}
  \label{fig:coag3d_turb_Machvar_expfits}
\end{figure}

Fig.~\ref{fig:coag3d_turb_Evary} and Fig.~\ref{fig:coag3d_turb_Evary_rseed} show the impact of the driving energy on the cold gas mass growth (and destruction). Droplets of size $r_{\rm d}\sim 5\lshatter$ still do not survive (except in the driving with $\mathcal{M}\sim 0.03$ which is essentially static) but the ones of size $500 \lshatter$ do survive with a growth rate dependent on the strength of the turbulence (see below).

We have seen that droplets of size $500\lshatter$ do survive in a turbulent medium with $\mathcal{M}\lesssim 1$ whereas droplets of size $5\lshatter$ do not -- even if they are grouped in large quantities ($N_{\rm d}\lesssim 1000$). Fig.~\ref{fig:coag3d_turb_rd50} shows now the mass evolution of droplets intermediate in size between those two, i.e., $r_{\rm d}\sim 50\lshatter$. One can see that runs with $\mathcal{M}\sim 1$ (green lines in Fig.~\ref{fig:coag3d_turb_rd50}) lead to a disintegration of the cold clouds whereas in runs with $\mathcal{M}\sim 0.1$ (shown in orange and red) the droplets survive and grow. The runs with $\mathcal{M}\sim 0.3$ do sometimes show growth and sometimes not. While the overall convergence appears somewhat better than in the $r_{\rm d}\sim 500\lshatter$ runs (cf. Fig.~\ref{fig:convergence_multi}), note that the survival scale $r_{\rm crit}$ is actually somewhat smaller for the $\mathcal{M}\sim 0.1, r \sim 50 \lshatter$ runs compared to the $\mathcal{M}\sim 1, r \sim 500 \lshatter$ runs (due to additional Mach number dependence in $r_{\rm crit}$; cf. \S~\ref{sec:disc_scale}).
These runs are closer to the $r_{\rm crit}$ threshold and worse convergence properties are reasonable.
The gray dotted line shows the expected mass growth rate of our model which again seems to fit reasonably well.

We carry out several  exponential fits to the $m(t)$ curves shown in Fig.~\ref{fig:coag3d_turb_Evary_rseed}. As indicated in the legend of Fig.~\ref{fig:coag3d_turb_Machvar_expfits}, we show the growth time for the full time ranges,
for $t<1.2 t_{\rm eddy}$ and $t > 1.2 t_{\rm eddy}$ (`early' and `late', respectively) -- but excluding the time when the mass growth flattens due to the limited simulation domain size, i.e., only for the time the cold gas mass fraction is $<30\%$. Fig.~\ref{fig:coag3d_turb_Machvar_expfits} shows the result of this fitting procedure where we normalize the growth time by Eq.~\eqref{eq:tgrow_TML_full} for $\mathcal{M}=1$ and set $l\sim r_\cl$ as before. We see that the measured growth times are a factor of $\sim 2$ too small compared to our analytic estimate (the dotted line follows our model with a fudge factor of $0.5$) and the Mach scaling is well reproduced.

\begin{figure}
  \centering
  \includegraphics[width=\linewidth]{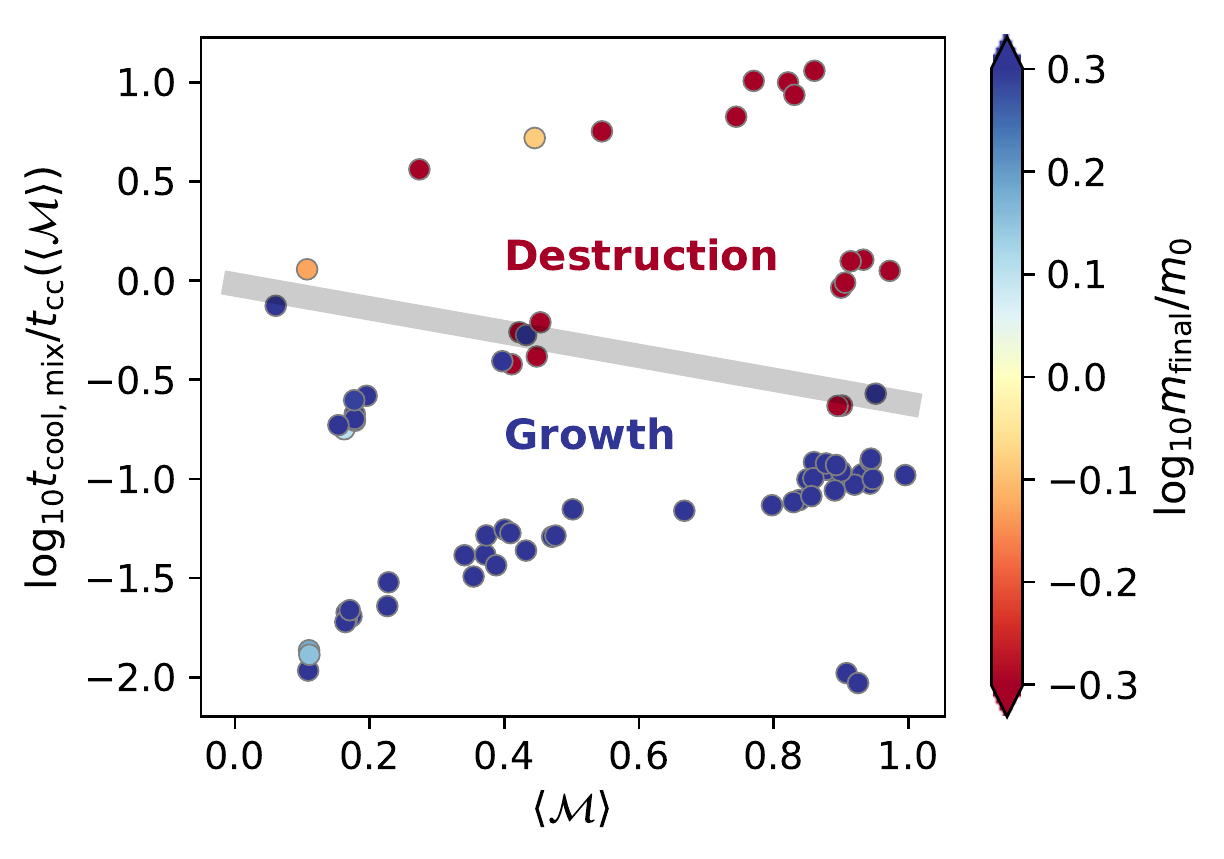}
  \caption{Overview of the simulations with $\chi\sim 100$. Shown is the measured rms Mach number with respect to the hot gas sound speed and the ratio of the cooling time of the mixed gas and the `cloud crushing' timescale. The color coding represents the final cold gas mass. Points were randomly offset in $y$-direction by $\pm 0.1$ for visualization purposes, so that points do not land on top of one another.}
  \label{fig:overview_trat_vs_mach}
\end{figure}

Figure~\ref{fig:overview_trat_vs_mach} is an overview of the previously discussed runs. Specifically, we show the measured Mach number $\mathcal{M}\equiv v_{\rm turb}/c_{\rm s,hot}$ (where we measured $v_{\rm turb}$ after it reached an approximate equilibrium) and the ratio of the cooling time of the mixed gas $t_{\rm cool,mix}$ with mixed gas temperature $T_{\rm mix} \sim (T_c T_h)^{1/2}$ being the geometric mean of the hot and the cold temperature and the `cloud crushing time' $t_{\rm cc}\sim \chi^{1/2} r_{\rm d} / v_{\rm turb}$\citep[as in][]{Gronke2018,Gronke2019}. The color coding represents the final mass with red and blue circles showing cold gas destruction and growth, respectively.
Fig.~\ref{fig:overview_trat_vs_mach} indicates that survival requires $t_{\rm cool,mix} / t_{\rm cc}\lesssim \alpha$ with $\alpha$ having a weak Mach number dependence (with $\alpha^{-1}\sim 3$ for $\mathcal{M}\sim 1$). This is nearly identical to the criterion found in `wind tunnel' simulations \citep{Gronke2018,Gronke2019,Kanjilal2020,Abruzzo2021,Farber2021} where, however, $\alpha\sim 1$.
We discuss this further in \S~\ref{sec:detail_survival} and \S~\ref{sec:disc_scale}.

\subsection{Effect of the overdensity $\chi$}
\label{sec:results_chi}

\begin{figure}
  \centering
  \includegraphics[width=\linewidth]{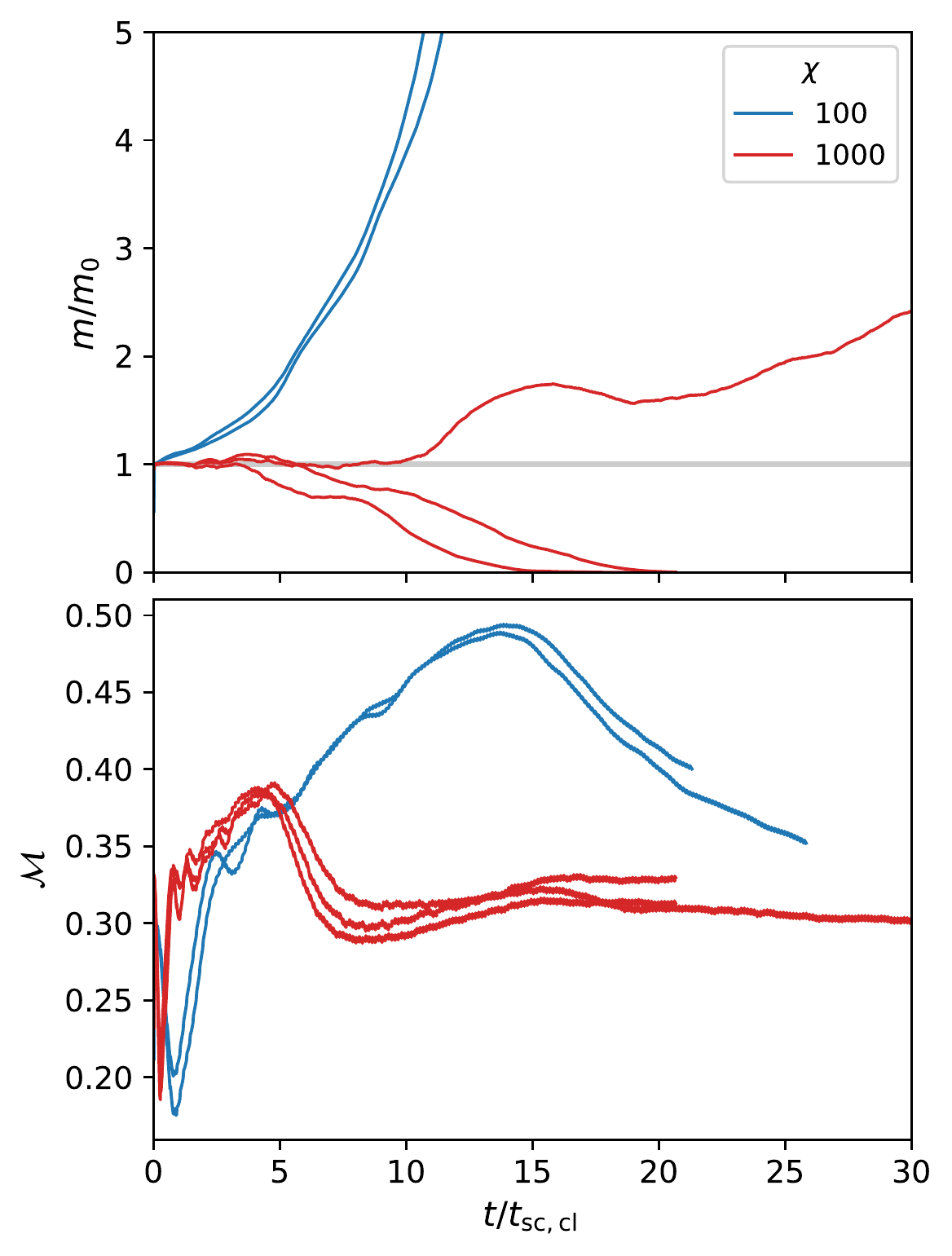}
  \caption{Mass and Mach number evolution of runs with size $r_\cl \sim 15 r_{\rm crit,w}$ but different overdensities. While the cold gas for $\chi\sim 100$ survives, survival becomes marginal at higher overdensities. As before, the same colors correspond to simulations with merely a different random seed as initial setup.}
  \label{fig:chi_rd500}
\end{figure}

\begin{figure}
  \centering
  \includegraphics[width=\linewidth]{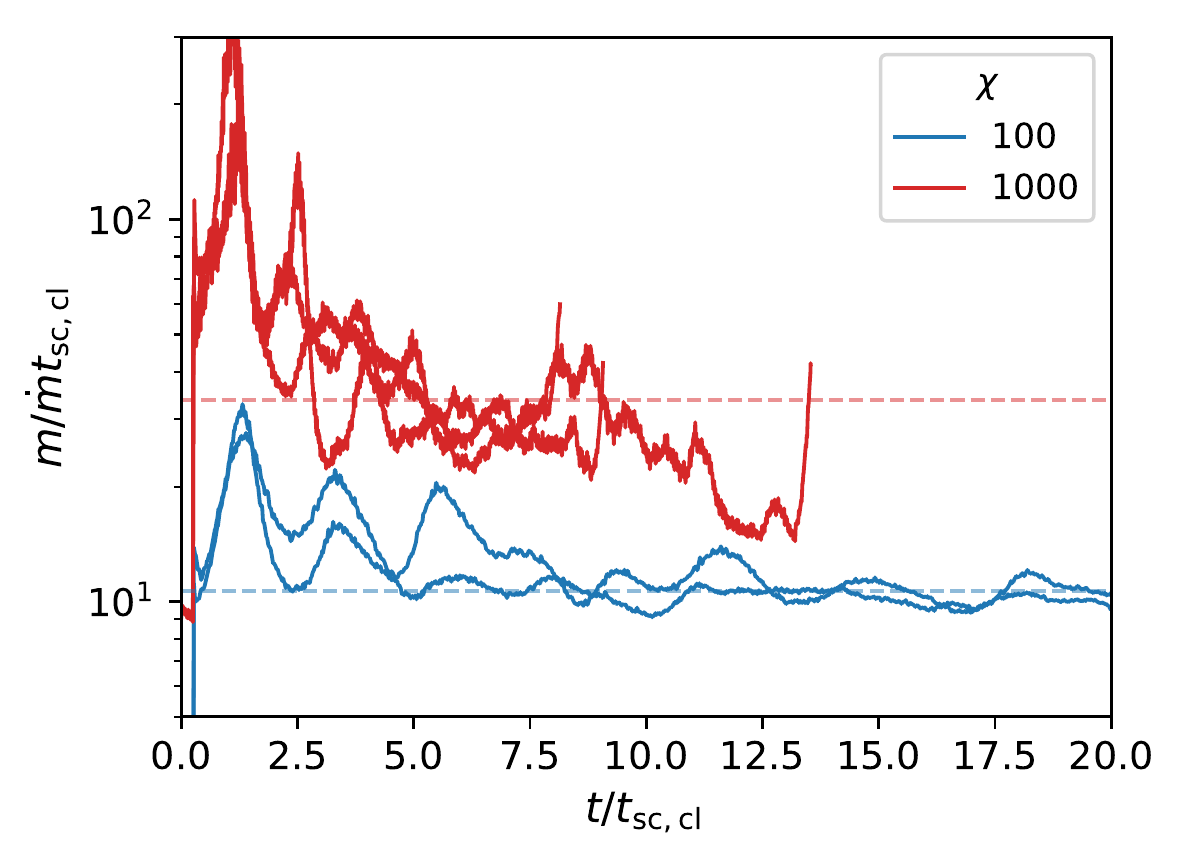}
  \caption{Mass growth time of runs with $\mathcal{M}\sim 0.1$ and size $r_\cl \sim 500 \lshatter$ ($r_\cl \sim 5000 \lshatter$) for $\chi\sim 100$ ($\chi\sim 1000$). The horizontal lines are the theoretical expectations using a fudge factor of $0.5$.}
  \label{fig:chi_rd500_M0.1}
\end{figure}

\begin{figure}
  \centering
  \includegraphics[width=\linewidth]{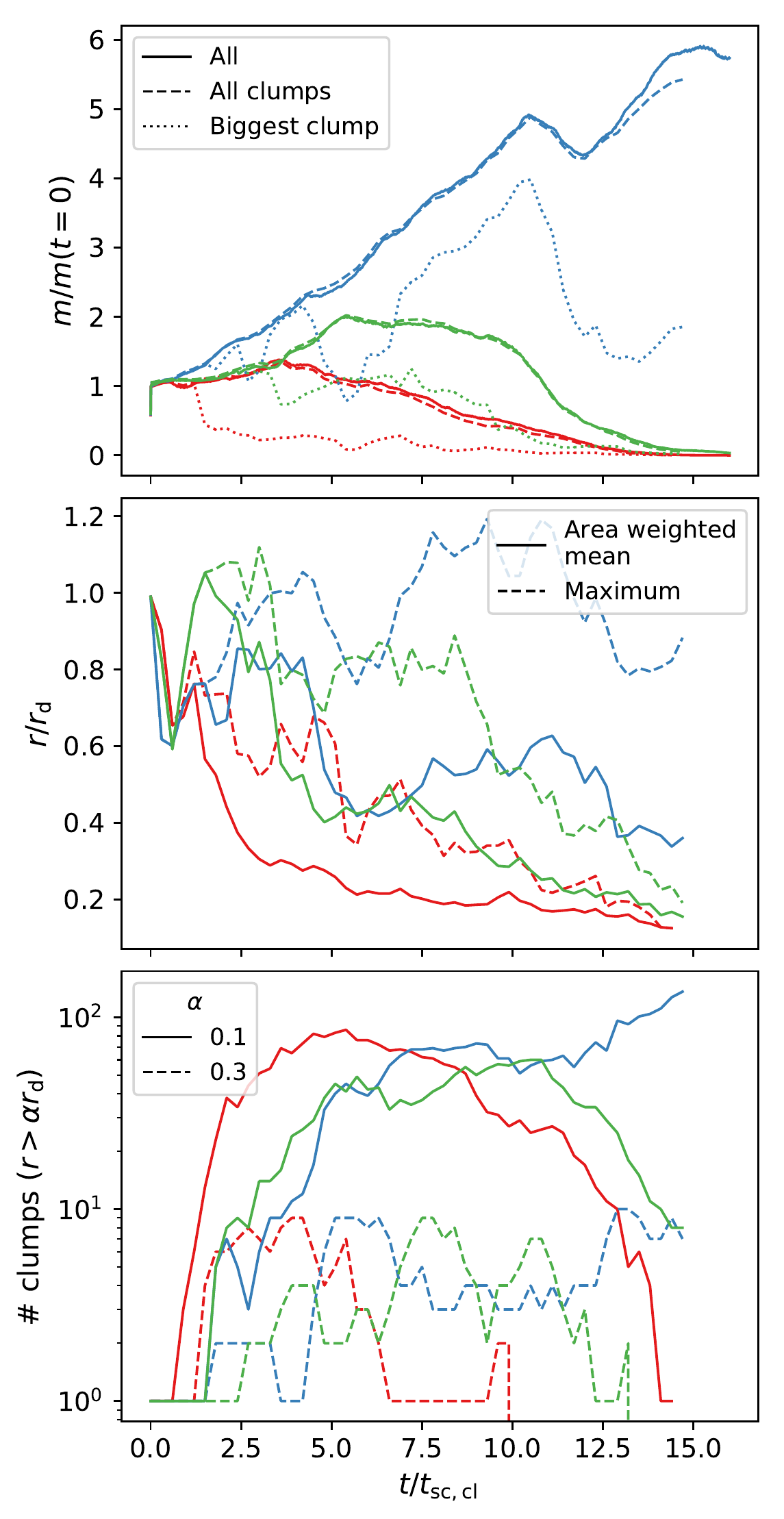}
  \caption{Evolution of three simulations of one droplet of size $r_{\rm d}\sim 200\lshatter$ placed in a $\mathcal{M}\sim 1$ turbulent medium with different random seed. The panels show (from top to bottom) the cold gas mass, the droplet size and numbers. Note while the run shown as the red curve has many more droplets, it lacks sufficiently big droplets to survive the turbulence.}
  \label{fig:multiplot_clump_analysis}
\end{figure}

Figure~\ref{fig:chi_rd500} shows runs with similar $t_{\rm cool,mix} / \tcc \sim 0.06$ for different overdensities. Cold gas in the $\chi\sim 100$ runs  survives (in fact, they do even for larger values of $t_{\rm cool,mix} / \tcc$, cf. Fig.~\ref{fig:overview_trat_vs_mach}) whereas cold gas survives in only one of the $\chi\sim 1000$ runs -- barely. At face value, this shift of the survival criterion may seem more aligned with the \citet{Li2019a,Sparre2020} criterion used for `windtunnel' simulations which compares the cooling time of the hot gas $t_{\rm cool,hot}$ to an empirically calibrated survival time $t_{\rm life}\sim \text{a few}\times \tcc$. The discrepancy between these two criteria worsens with larger overdensities (cf. figure 1 in \citealp{Kanjilal2020}).
Several subsequent studies have compared the two criteria put forward and suggested that $t_{\rm cool,mix}/\tcc \sim 1$ fits the cold gas survival in a wind tunnel better -- if one follows the non-monotonic mass evolution -- which dips and then recovers -- out to late times \citep{Kanjilal2020,Abruzzo2021}. This is because, for these large overdensities, the cloud first loses a substantial amount of cold gas before being `reborn' from the mixed medium (\citealp[][]{Farber2021} find in their lower temperature clouds that this transition occurs when the mixing timescale within the cloud is comparable to the cooling time). Hence, if one cuts off this regrowing phase, one might conclude that the final cold gas mass is lower than the initial one. This evolution for a laminar flow is relevant for the observed behavior in this study: since this `regrowing phase' occurs when cold gas is entrained, this is less likely to occur in a turbulent setup when the gas velocity is continually changing in direction and magnitude. It is clearly affected by driving properties such as the correlation time $t_{\rm corr}$\footnote{Numerical resolution can also play a role. We ran simulations with $1/8$ the mass resolution and found cold gas destruction rates to increase.}. These runs are close to the survival threshold (as can be seen from Fig \ref{fig:overview_trat_vs_mach}, $\alpha \sim 0.5$ at these turbulent Mach numbers, thus $r_{\rm cl} \sim 15 r_{\rm crit, w} \sim 8 r_{\rm crit, turb})$. Besides the Mach number dependence (as demonstrated in Fig  \ref{fig:overview_trat_vs_mach}), we expect the dimensionless parameter $\alpha$ in Eq.~\eqref{eq:tcoolmix_crit} to have some additional overdensity dependence as well, which could account for the behavior seen in Fig. \ref{fig:chi_rd500}. We will discuss additional contributing factors (non-constant $t_{\rm grow}$, difference in Mach numbers) below. While the  additional overdensity dependence of $\alpha$ is interesting, in the remainder of this paper we focus on the $\chi\sim 100$ regime, which is most relevant for the CGM of $L_*$ galaxies.

That a cloud with $\chi\sim 1000$ can easily survive is shown in Fig.~\ref{fig:chi_rd500_M0.1}. Here, we show results from simulations with $\chi=(100,\,1000)$, $\mathcal{M}\sim 0.1$ (i.e., with a lower Mach number than in Fig.~\ref{fig:chi_rd500}).
Specifically, Fig.~\ref{fig:chi_rd500_M0.1} shows the growth time of the cold gas defined as $t_{\rm grow}= m/\dot m$ with the horizontal lines corresponding to the analytic values of Eq.~\eqref{eq:tgrow_TML_full} (using a fixed fudge factor of $0.5$ and $l=r_{\rm cl}$). Note that the simulations shown are with a constant physical cloud size, i.e., we expect $t_{\rm grow}/t_{\rm sc,cl}\propto \chi\lshatter^{-1/2}c_{\rm s}/r_\cl\propto \chi^{1/2}$ which agrees with the numerical findings. Fig.~\ref{fig:chi_rd500_M0.1} hints that the  growth time in the $\chi=1000$ cases decreases after an initial phase. If so, this would be consistent with similar behavior in wind-tunnel simulations, where growth kicks in at late times. It is unclear what drives such behavior here, though fragmentation into smaller droplets (and a decrease in the length scale $l$ in equation \ref{eq:tgrow_TML_full}) is a possibility.

\subsection{A closer look at droplet survival}
\label{sec:detail_survival}

\begin{figure}
  \centering
  \includegraphics[width=\linewidth]{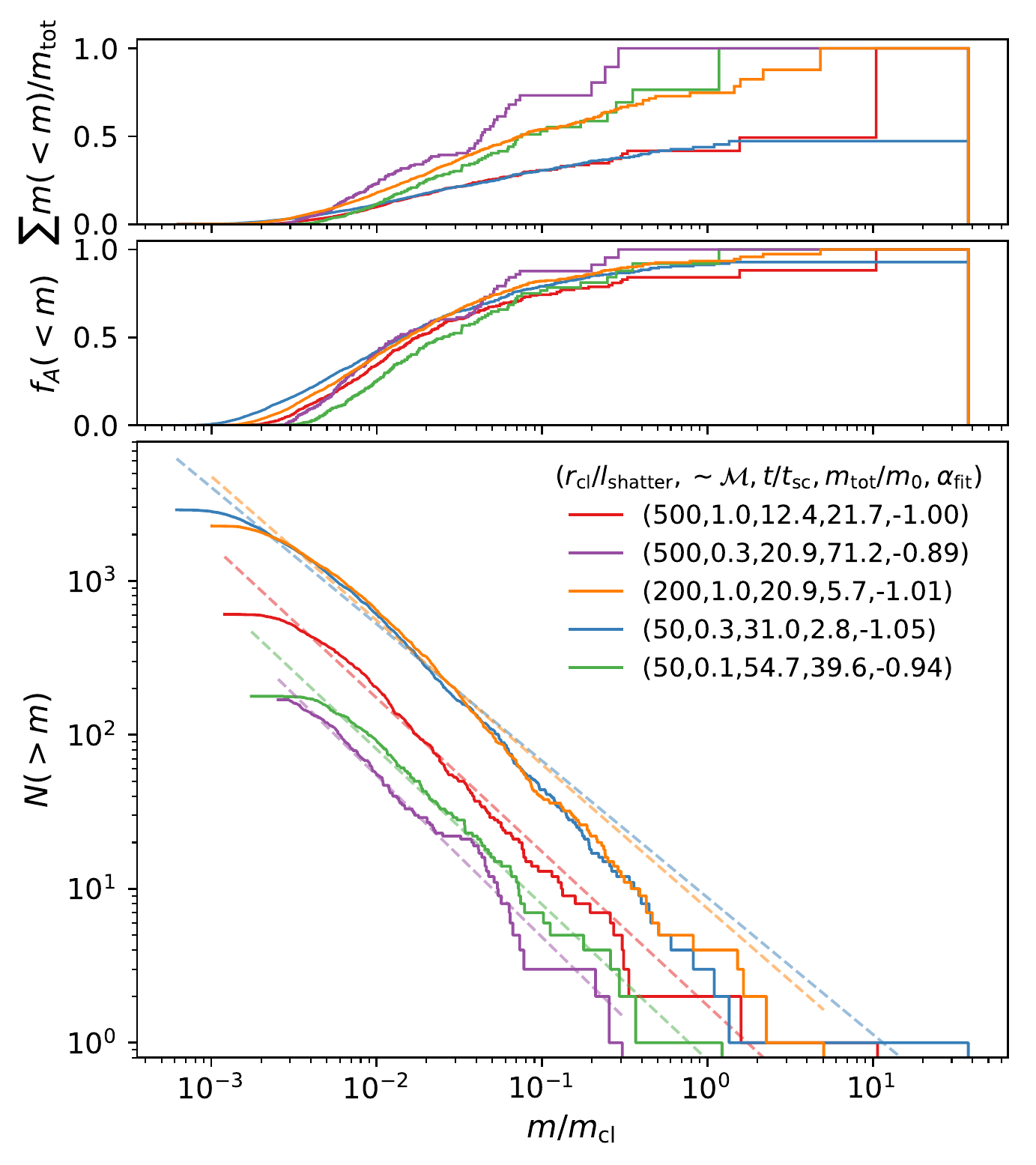}
  \caption{Cumulative droplet mass distribution with power law fits (bottom panel), and cumulative distributions weighted by mass and projected area (top and central panel, respectively). While the mass is $\gtrsim 50\%$ in droplets smaller than the initial cloud size, the area is fully dominated by these smaller droplets.
The simulations shown are the same as in Fig.~\ref{fig:clump_analysis_multiplot_severalsim} using the same color coding.
  }
  \label{fig:mcumhist}
\end{figure}

\begin{figure}
  \centering
  \includegraphics[width=\linewidth]{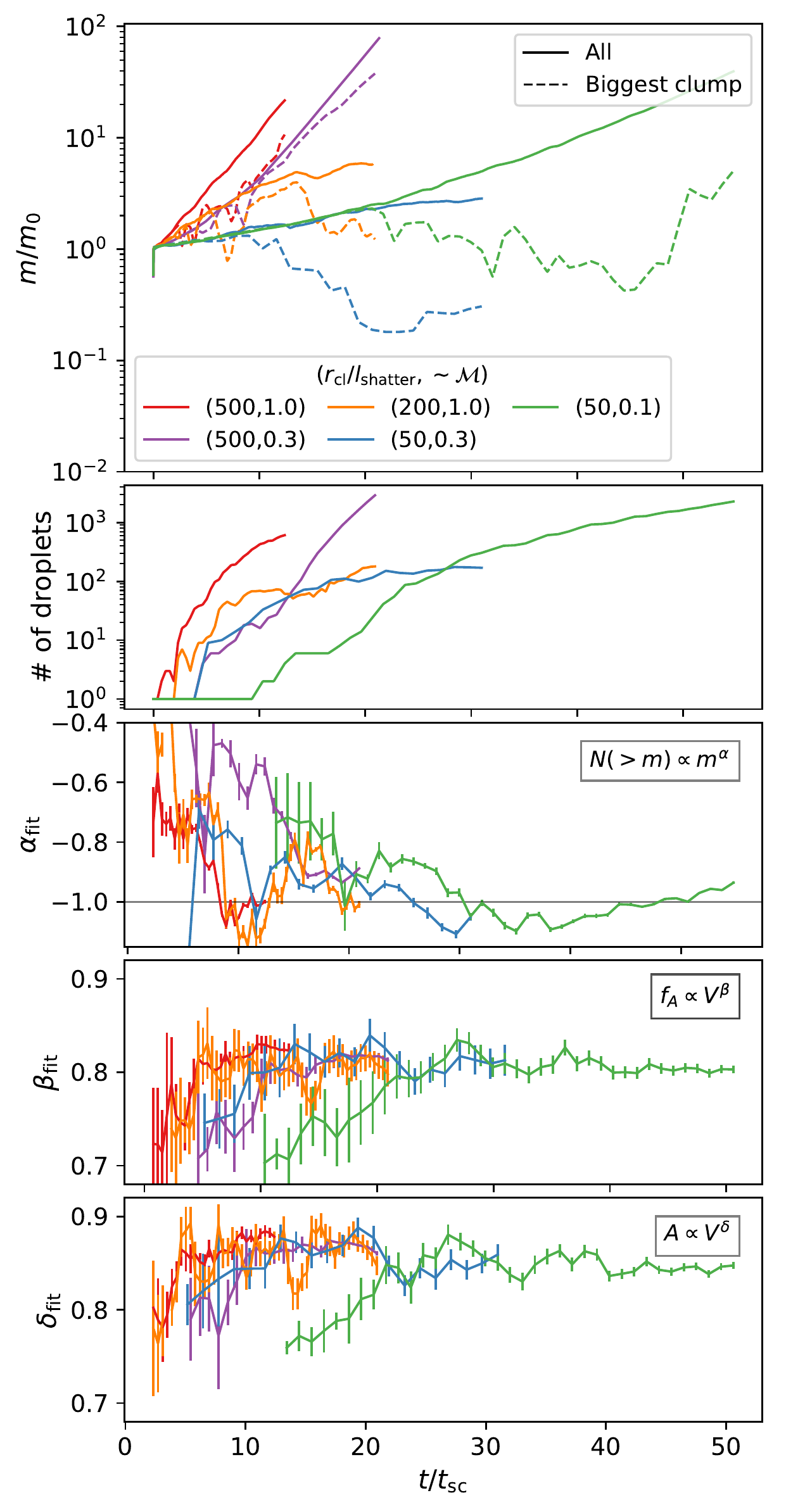}\\
  \caption{Evolution of cold gas mass, number of identified droplets, and power-law exponents of the projected area-volume and surface area-volume correlations and cumulative mass distribution (from top to bottom).
    The simulations shown are examples with different initial cloud sizes and Mach numbers. However, at later times all the mass distributions tend to follow $\dd N / \dd m\propto m^{\alpha-1}$ with $\alpha\sim -1$ (see Fig.~\ref{fig:mcumhist} for distributions of the simulations shown here at a single snapshot).
}
  \label{fig:clump_analysis_multiplot_severalsim}
\end{figure}

\begin{figure}
  \centering
  \includegraphics[width=\linewidth]{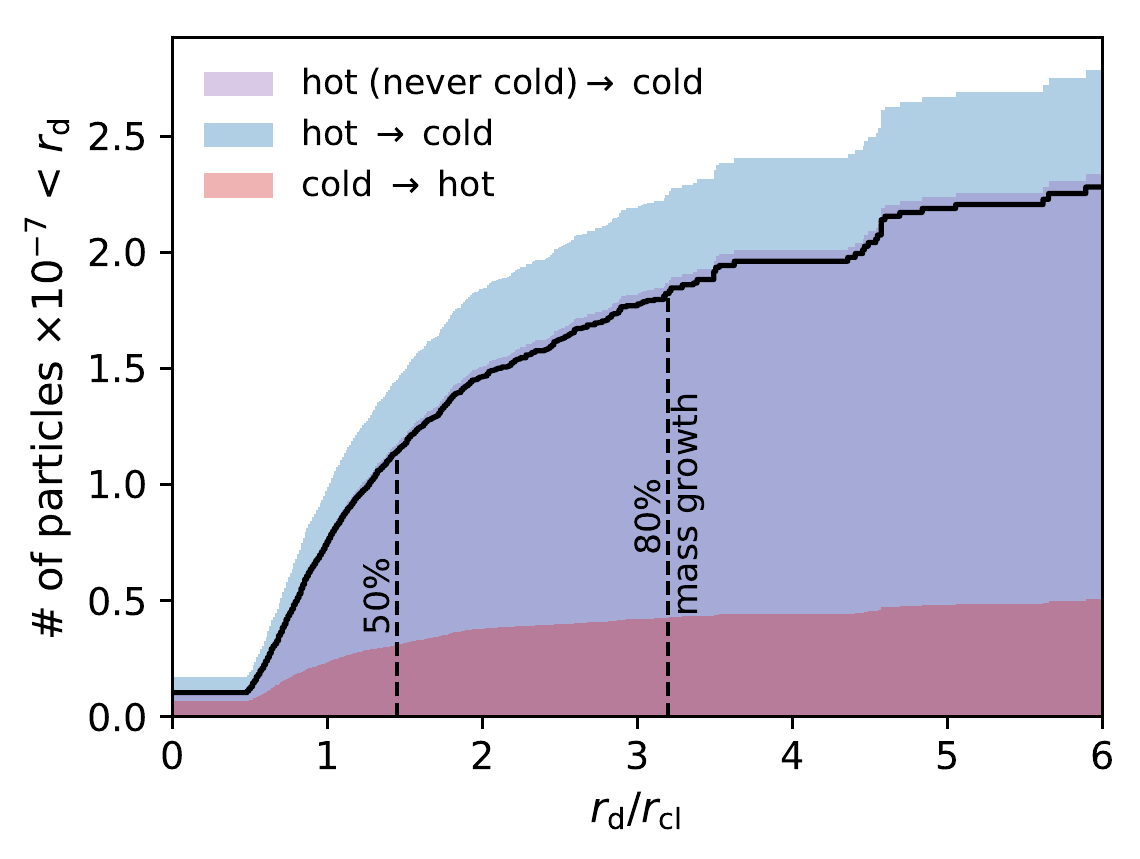}
  \caption{Cumulative distribution function of tracer particles moving from hot to cold gas and vice-versa. The black solid line shows the corresponding mass growth (i.e., the hot$\rightarrow\,$cold transitions minus the cold $\rightarrow\,$hot ones). The simulation shown is with $\mathcal{M}\sim 0.4$ detailed in \S~\ref{sec:droplet_dists}.}
  \label{fig:fig_dmparts_cumhist}
\end{figure}

\begin{figure}
  \centering
  \includegraphics[width=\linewidth]{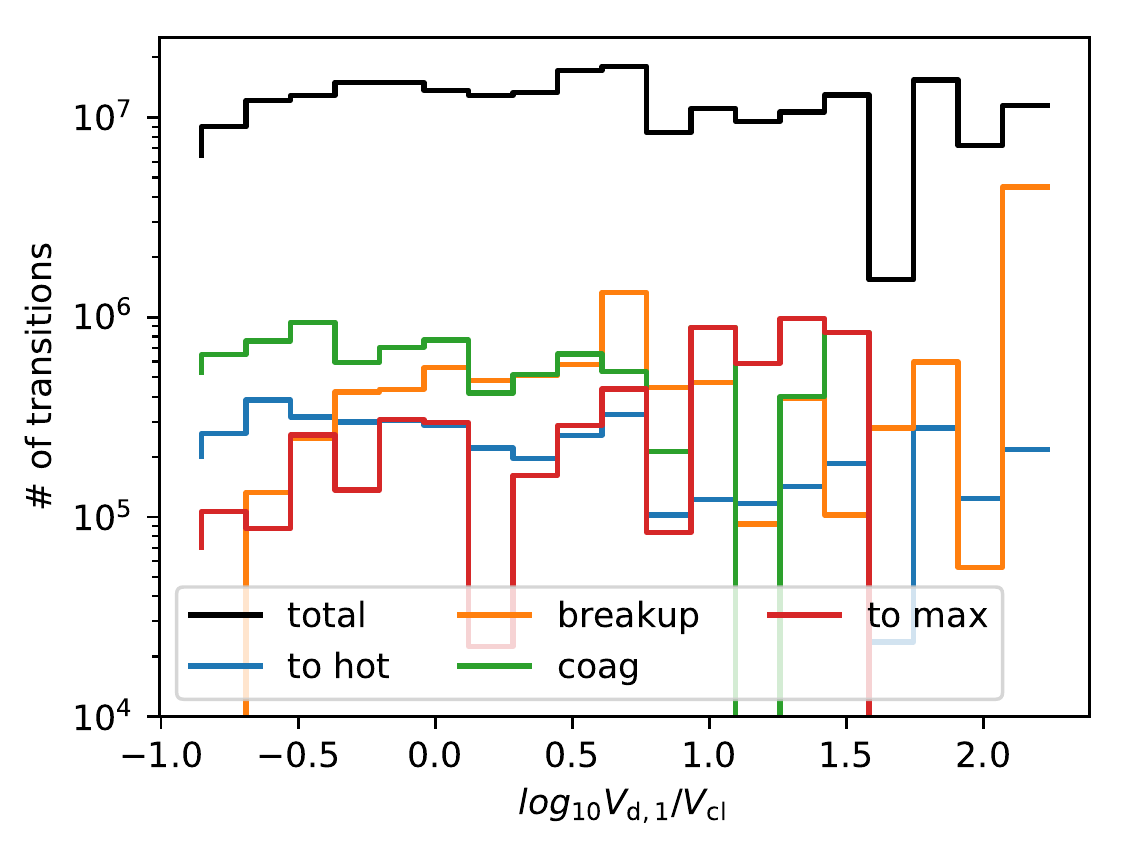}
  \caption{Breakdown of cumulative transitions of tracer particles as a function of the clump volume they belong to prior to the transition. Here, `breakups' and `coagulations' are defined to transitions below $50\%$ and above twice the prior size, respectively. The `to max' line indicates transitions to the biggest clump mass. Note how most of the transitions do  not fall in any of these categories, i.e., represent `natural' growth / mass loss -- independent of the clump size.
  }
  \label{fig:transitions_vs_size}
\end{figure}

\begin{figure}
  \centering
  \includegraphics[width=0.9\linewidth]{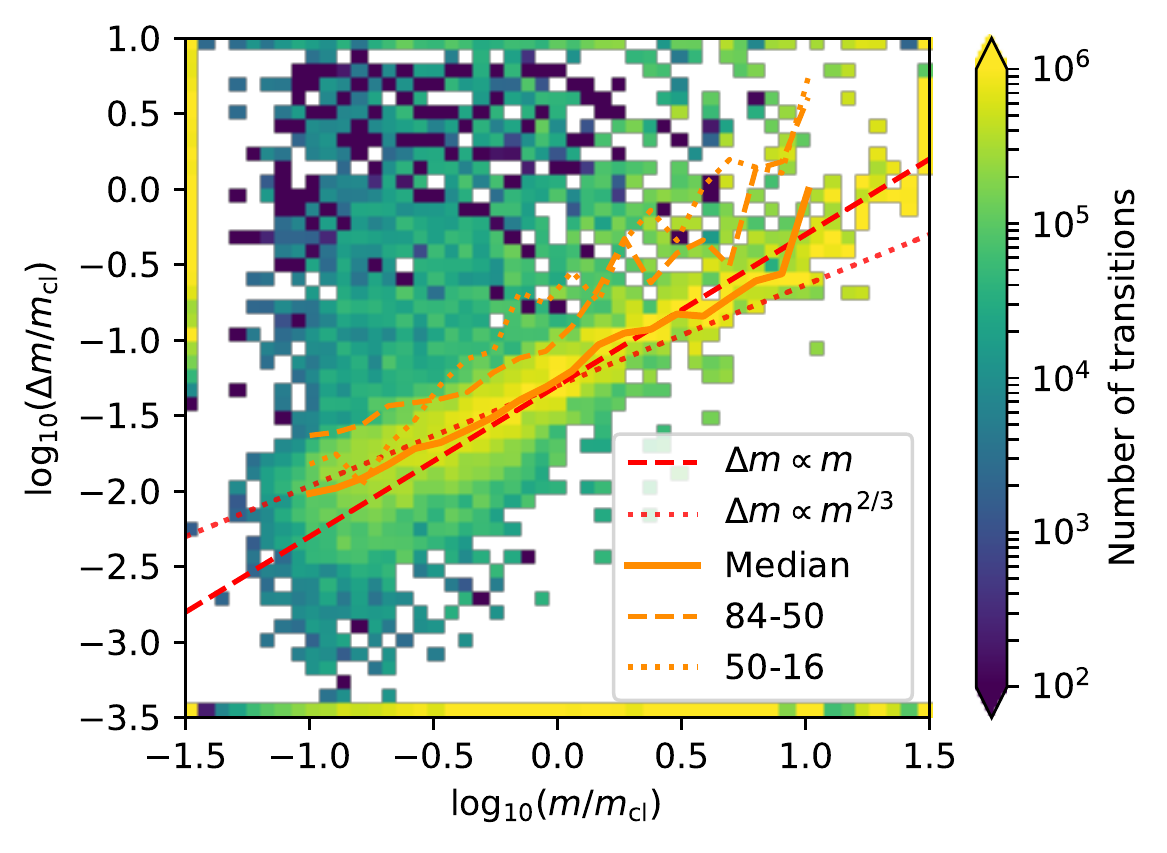}
  \caption{Distribution of clump growth as tracked by tracer particles with the particles falling outside the shown range clipped to be visible at the boundary. The red line shows the $\dot m\propto m$ relation for visual aid, and the orange line shows the median as well as the difference between the median and the $84$th/$16$th percentiles. Most transitions fall near that relation indicating clumps growing `naturally'. However, there are a significant number of transitions above the line and at the bottom of the plot showing coagulation and fragmentation, respectively. See Fig.~\ref{fig:fig_dmparts_cumhist} for details on the phase transitions.}
  \label{fig:hist2d_dm_vs_m_clumps}
\end{figure}

Figure~\ref{fig:multiplot_clump_analysis} shows in more detail a run with $r_{\rm d}\sim 200\lshatter$ and $\mathcal{M}\sim 1$ with the different curves corresponding to different random seeds. From Eq.~\eqref{eq:rcrit_lshatter}, and assuming $\alpha \sim 0.3$ from Fig.~\ref{fig:overview_trat_vs_mach} at $\mathcal{M} \sim 1$, this is close to the threshold for survival $r \sim 1.3 r_{\rm crit}$. This can be seen from the top panel of Fig.~\ref{fig:multiplot_clump_analysis}, which shows the cold mass evolution. As we have seen before (cf. Fig.~\ref{fig:multi2d_3dturb_N1_rdvar}), due to the turbulence the droplets break up into smaller clumps. The lower two panels of Fig.~\ref{fig:multiplot_clump_analysis} show the evolution of the size of the droplets\footnote{We identified the droplets as a region of minimum $16$ cells with $T < 2 T_{\mathrm{floor}}$ not touching another such region. Specifically, we used \texttt{ndimage.measurements.label} of the \texttt{Python} package \texttt{scipy} \citep{scipy}.}.
In only one of the runs shown in Fig.~\ref{fig:multiplot_clump_analysis}, the cold gas survives. Specifically, this is the case for the run in which a large cloud survives and is not broken up into many droplets. This intuitively makes sense, as a breakup can lead to a scenario where the fragments are too small to fulfill the survival criterion Eq.~\eqref{eq:tcoolmix_crit}. We discuss this further in \S~\ref{sec:CGM_origin}. These results also highlight the stochasticity of outcomes near the survival boundary, which we discuss further in \S~\ref{sec:results_stochastic}. In general, as one nears the survival boundary, stochasticity and resolution effects play an increasingly important role. 

\subsection{The droplet growth and mass distribution}
\label{sec:droplet_dists}
We can study the droplet growth and mass distribution resulting from the fragmentation process a bit further. Figure~\ref{fig:mcumhist} shows the mass distribution of droplets at the final stage of different simulations which show mass growth (lower plot) alongside the fractional distributions of mass and areal covering fraction (in the upper two panels). 
In spite of the differences between the simulations, the cumulative mass distributions follow $N(>m)\propto m^{-\alpha}$ with $\alpha\sim -1$ as also shown by the power-law fits (the resulting exponent $\alpha_{\rm fit}$ is shown in the legend of the lower panel).
The normalized mass and covering fraction distributions show that while the mass can be dominated by some larger clumps, the area covering fraction is dominated by small clumps. This has observational implications which we discuss in \S~\ref{sec:CGM_cold_gas_shape}.

Figure~\ref{fig:clump_analysis_multiplot_severalsim} shows the evolution of these simulations (with the same color coding as Fig.~\ref{fig:mcumhist}). Specifically, in the top part of Fig.~\ref{fig:clump_analysis_multiplot_severalsim}, we show (from top to bottom) the cold gas mass evolution, the number of identified droplets, and the power-law index from fitting the cumulative clump mass distribution at various snapshots. From the latter, it is visible that the distributions start out flatter but then converge to $\alpha\sim -1$ after some time (and a sufficient number of droplets).

The lower two rows of Fig.~\ref{fig:clump_analysis_multiplot_severalsim} show the relation of the areal covering factor of the droplets and the surface area (measured as described above) with the droplets' volumes.  We find the relations to be steeper than for pure monolithic growth, indicative of a corrugated surface area and thus larger mass growth not only for the cold gas as a whole but also on a per-droplet basis. 
As discussed in \S\ref{sec:analytics}, steeper than Euclidean $(A \propto V^{2/3}$) scaling is expected for fractal geometry. Indeed, $A \propto V^{\delta}$ where $\delta=D/3=0.83$ fits our measured $\delta=0.8$ well, if the fractal dimension $D\approx 2.5$, as measured in numerical simulations and experiments of mixing layers \citep{Sreenivasan1986,Fielding2020}\footnote{\citet{Sreenivasan1989} derived $D=7/3$ which they found in excellent agreement with experimental data \citep{Sreenivasan1986}.}.
In order to investigate this further and to  obtain the mass growth distribution as a function of droplet size, we resort to Lagrangian tracer particles which we inject uniformly in our \texttt{FLASH} simulation (see \S~\ref{sec:methods} for detailed numerical setup).

Figure~\ref{fig:fig_dmparts_cumhist} shows the number of tracer particles that change from hot to cold medium or the other way around as a function of droplet size. The non-zero offset at $r_{\rm d}=0$ is due to cold material in clumps smaller than our clump detection threshold. We can see that the mass growth, i.e., the number of particles changing from hot to cold medium minus the ones changing from cold to hot (marked with the black solid line), has approximate equal contributions from clumps $\lesssim 2 r_{\rm cl}$ in size which in total already contribute $> 50\%$ of the total mass growth (which is $>300 m_{0}$ for the simulation shown). Or in other words: the mass growth is not dominated by few large clumps but instead by many smaller ones. 
Note that the net mass growth follows the supply of fresh hot gas (shown as the black line and the filled red area in Fig.~\ref{fig:fig_dmparts_cumhist}, respectively), which must be true in steady state, since recycled (i.e., that was heated up and cooled back down) gas does not contribute to net cold gas mass growth. 

The impact of `breakups' and `coagulations' for this simulations is shown in Fig.~\ref{fig:transitions_vs_size}. Here, we show the time integrated transitions split up into ``to hot'', i.e., to the hot medium, ``breakups'' and ``coagulations'' where the droplet size after the transition is $0.5$ and $2$ times the prior one, respectively. This figure shows that the fractional impact of breakups and coagulations does not depend on the clump size and is overall small ($\lesssim 10\%$) compared to the `natural' growth or mass loss (in line with Fig.~\ref{fig:fig_dmparts_cumhist} showing that mass growth occurs over a range of droplet sizes).
In Fig.~\ref{fig:transitions_vs_size} we also mark the transitions to the maximum clump mass at a given time, i.e., coagulations of small clumps to the biggest clump. As we will show in \S~\ref{sec:mass_dist} these transitions are important for the formation of the $\dd N(m)/\dd \log m \propto m^{-1}$
clump mass distribution.

We show details on all the transitions that end up or start within the cold medium in Fig.~\ref{fig:hist2d_dm_vs_m_clumps} (for the same simulation and time range as shown in Fig.~\ref{fig:fig_dmparts_cumhist}). It shows the distribution of clump mass differences between two snapshots ($\Delta t\sim 0.3 t_{\rm sc,cl}$) versus the initial clump masses. One can clearly see that most of the transitions follow the $\dot m\propto m$ relation but a number of transitions fall below and above that relation showing fragmentation and coagulation, respectively. While the former results in clumps with masses of one over $\sim$ a few times the original mass, in the latter process clumps merge commonly with others that are $\gtrsim$ an order of magnitude in mass. Figure~\ref{fig:hist2d_dm_vs_m_clumps} shows that for larger clumps, $\dot m\propto m$ and the scatter around this relation follows an approximate linear relationship.
The median relationship flattens somewhat for lower mass clumps. However, lower mass clumps are heavily affected by coagulation, which may be driving this. Resolution effects could also play a role.

\subsection{Stochasticity of the evolution and convergence}
\label{sec:results_stochastic}

\begin{figure*}
  \centering
  \includegraphics[width=\textwidth]{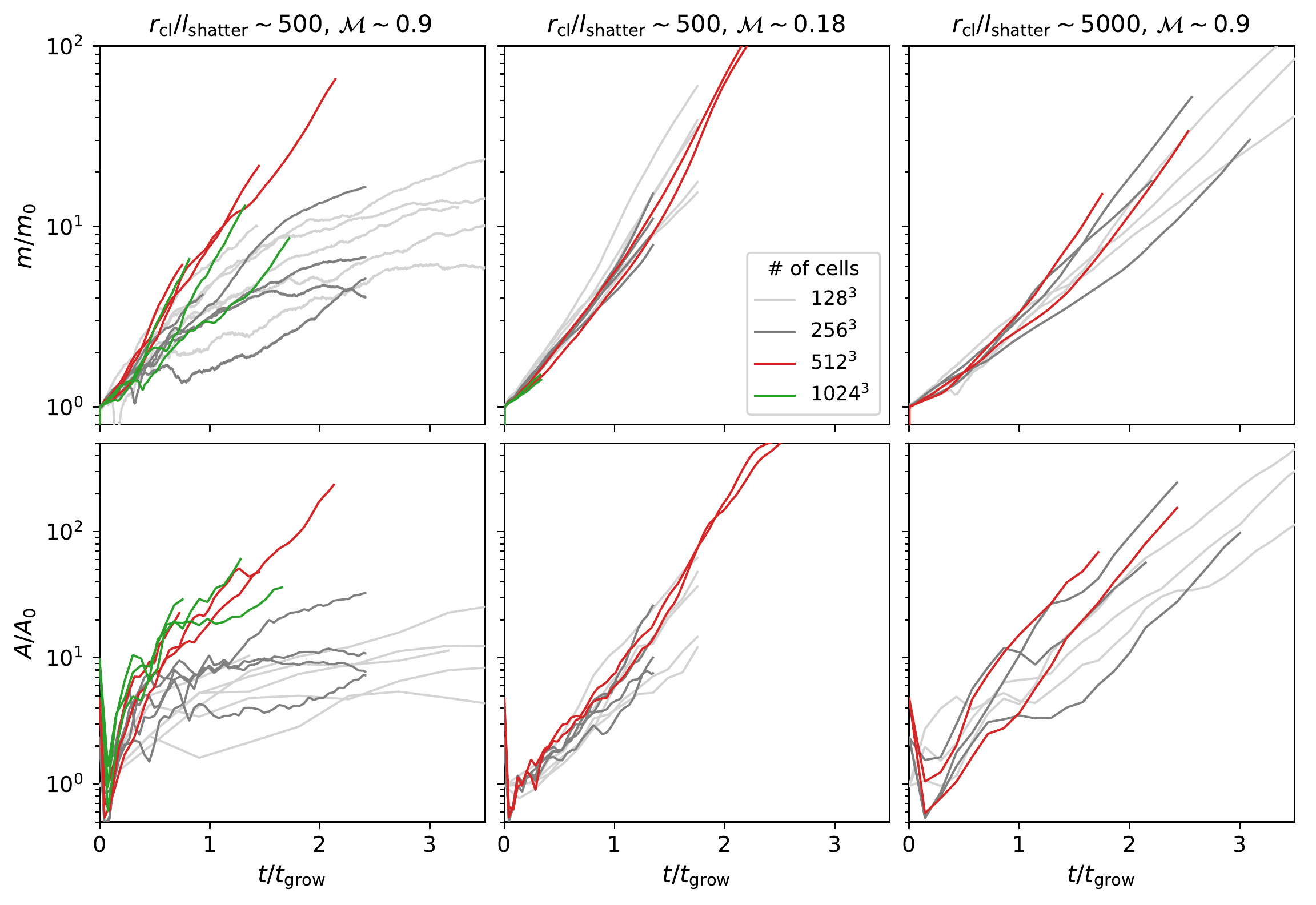}
  \caption{Mass and surface area evolution (top and bottom row, respectively) of a single cloud with different sizes, Mach numbers (as annotated in the title of each column) and resolutions. The several lines with the same color corresponds to the same setup with a different random seed.} 
  \label{fig:convergence_multi}
\end{figure*}

One important thing to consider is that the stirring and the turbulent nature of the problem make it highly stochastic, i.e., the outcome of the `same' simulation using a different random seed (i.e., a different stirring pattern) can lead to a very different outcome of the problem. This is illustrated in Fig.~\ref{fig:convergence_multi} where we show several realizations of our setup indicated by the same color. Specifically, the upper and lower row of Fig.~\ref{fig:convergence_multi} shows the cold gas mass and surface area evolution, respectively.
 Note that the mass growth rate varies significantly between the runs. This means that studying the effect of, for instance, the turbulent properties on the cold gas mass growth rate is only possible in a statistical sense
 \footnote{In reality, the large volume  of the multiphase gas will lead to a mean mass growth rate which we can try to emulate by averaging over several runs}.
 
The left column of Fig.~\ref{fig:convergence_multi} show considerable stochasticity and resolution dependence near the survival boundary, when $r\sim 3 r_{\rm crit}$. However, in the subsequent middle and right panels, where $r \gg r_{\rm crit}$, this stochasticity and resolution dependence is significantly reduced, even though $r_{\rm crit}$ is not resolved in these simulations. This makes sense: since the mass distribution is scale-free, with equal mass per logarithmic interval (cf. \S~\ref{sec:mass_dist}). It is not necessary to resolve $r_{\rm crit}$ to achieve convergence, as long as median cloud sizes (which is presumably seeded by some external process like thermal instability) are much larger than the cell size \textit{and} the critical scale $r_{\rm crit}$. If so, one is only logarithmically sensitive to the cut-off scale.  If one is close to the critical scale $r_{\rm crit}$, then there is more much resolution dependence and sample variance depending on whether the largest cloud dips above or below $r_{\rm crit}$, and hence much more stochasticity (cf.~\S~\ref{sec:detail_survival}).

\subsection{Relation to observables}
\label{sec:res_observables}

Due to the domain size, the actual observables depend on our input parameters (cf. \S~\ref{sec:disc_caveats} for a discussion of the caveats). Here, we discuss some general scalings which might be useful in the comparison with observations.

\subsubsection{Cooling luminosity}
\label{sec:res_luminosity}

\begin{figure}
  \centering
  \includegraphics[width=\linewidth]{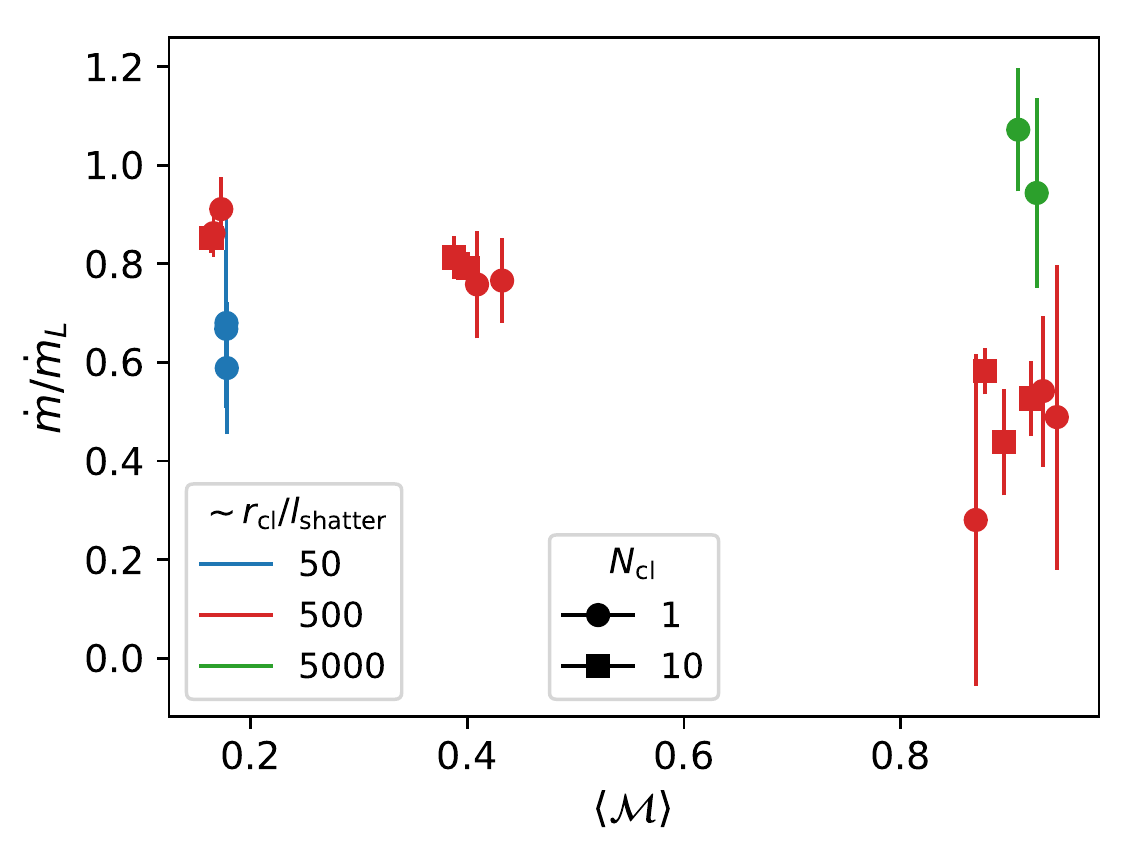}
  \caption{Ratio of the cold gas mass growth and the one expected from the recorded cooling emissivity (Eq.~\eqref{eq:mdot_L}) versus the measured rms Mach number. Points and error bars indicate the $16$th, $50$th and $84$th percentile of $\dot m / \dot m_{L}$ taken over $t/t_{\rm sc,cl}\in [2,\,20]$.}
  \label{fig:mdot_L_overview}
\end{figure}

Figure~\ref{fig:mdot_L_overview} compares the numerically obtained mass growth to the one expected from the recorded cooling emission, that is,
\begin{equation}
  \label{eq:mdot_L}
  \dot m_L = \frac{2}{5} \frac{\mu m_{\rm p}}{k_{\rm B} T_{\rm h}} \frac{L}{\mathcal{M}^2 + 1}
\end{equation}
where we used the rms velocity of the simulation domain to obtain the Mach number $\langle \mathcal{M}\rangle$. We see that for a variety of cloud sizes (as long as mass growth dominates) and Mach numbers $\dot m \sim \dot m_L$ holds. For the $r_\cl \sim 500 \lshatter$ runs, we find $\dot m_{L}$ slightly lower than $\dot m$ which might be due to larger turbulent dissipation. Note that the $(1+\mathcal{M}^2)$ term \citep[cf.][]{Ji2018}  is required in order to derive a mass growth rate from observed cooling emission.

\subsubsection{Velocity structure functions}
\label{sec:res_velocity}

\begin{figure}
  \centering
  \includegraphics[width=\linewidth]{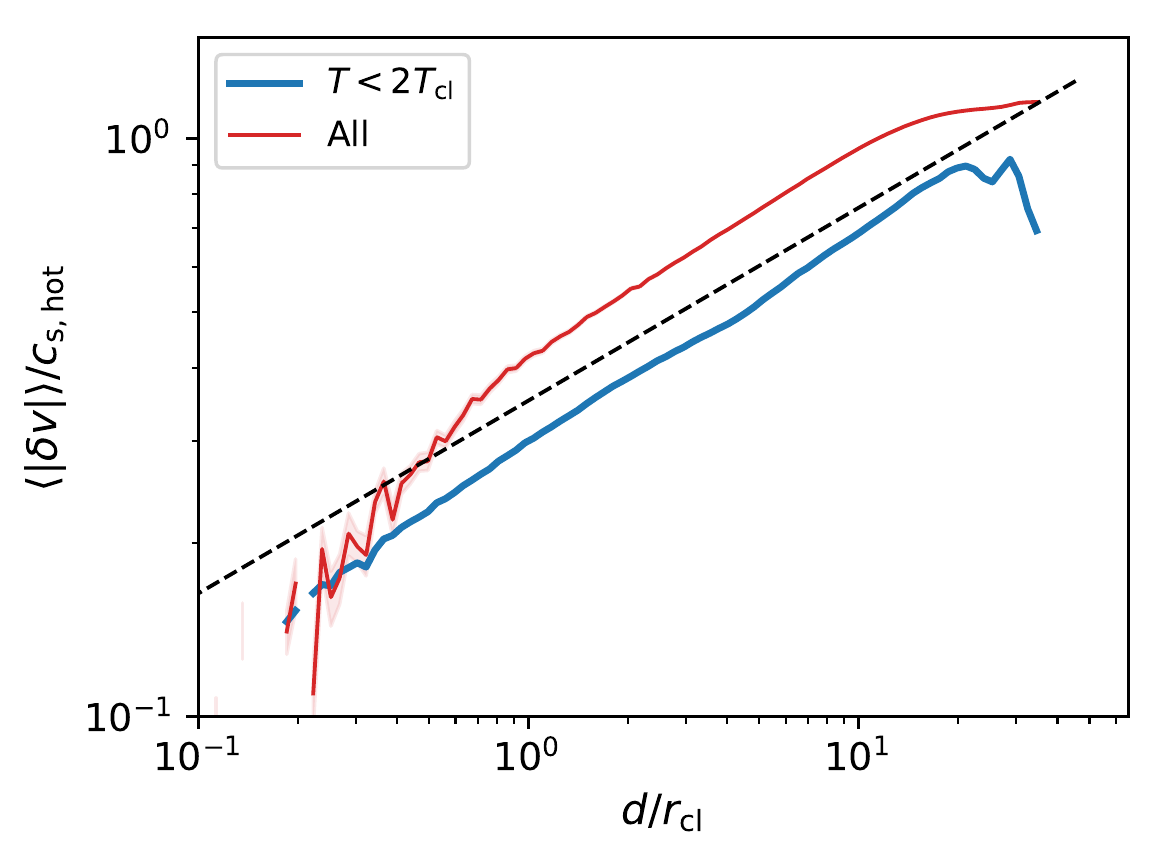}
  \caption{Representative example of a first order velocity structure function (shown is a simulation with $\mathcal{M}\sim 1$ and $r_\cl / \lshatter \sim 500$ at $t\sim 18 t_{\rm sc,cl}$). Black dashed line shows a $1/3$ expected from Kolmogorov turbulence.}
  \label{fig:vsf_rd500}
\end{figure}

Velocity structure functions (VSFs) show the mean velocity difference (to the power of the order of the VSF) as a function of distance and are commonly used in astrophysics to assess turbulent gas dynamics \citep[e.g.,][]{Chira2019,Ha2021}. Fig.~\ref{fig:vsf_rd500} shows the first order VSF for one of our $\mathcal{M}\sim 1$, $r_\cl/\lshatter \sim 500$ simulations, which is defined as $\langle \delta v \rangle(d) = \langle v(r+d) - v(r) \rangle$, averaged over all positions $r$.  Specifically, it shows the VSF for all material as well as for the cold ($T < 2 T_\cl$) at $t\sim 18 t_{\rm sc,cl}$, and to limit computational time using only $2\times 10^4$ randomly drawn cells. We can see that (i) the curves follow each other but the cold gas is at somewhat lower velocities, indicating imperfect entrainment, and (ii) they approximately have a $1/3$ slope expected from Kolmogorov subsonic turbulence. For small scales, the slope is steeper due to numerical dissipation; cf. \S~\ref{sec:disc_caveats} for a discussion of the caveats.
This behavior has implications for the observable line-widths in absorption spectra which we discuss in \S~\ref{sec:CGM_cold_gas_shape} and will revisit in future work. Overall, this shows that mass transfer from hot to cold gas via mixing and cooling can enforce good momentum coupling and entrainment between hot and cold phases (see, e.g., \citealp{Gronke2018} and \citealp{2021ApJ...911...68T,Schneider2020} for a discussion of this effect in `wind tunnel' and galactic simulations, respectively), as required if the more easily detected cold phase is to serve as a tracer of hot gas kinematics.

\section{Discussion}
\label{sec:discussion}

In this section, we discuss the implications of our findings for larger scale simulations (\S~\ref{sec:disc_scale}) and the circumgalactic medium (\S~\ref{sec:CGM_origin}, \S~\ref{sec:CGM_cold_gas_shape}). We also discuss the robust, universal $m^{-2}$ law for the mass distribution of droplets in \S~\ref{sec:mass_dist} and the caveats of this study in \S~\ref{sec:disc_caveats}.

\subsection{Convergence in turbulent, multiphase media}
\label{sec:disc_scale}
Converged simulations (of astrophysical, multiphase systems) are fundamental in order to compare them to observations (which are often based on the cold gas phase) or to make predictions that depend heavily on the cold gas (such as, e.g., statements about the ionizing photon escape).
It is, thus, logical to ask which scale one would need to resolve in order to obtain `first order' convergence, that is, in the total cold gas mass. 

One can rewrite the survival criterion found in this study, to a length scale criterion \citep[akin to what was done in][]{Gronke2018} which is for the here tested, i.e., $T_{\rm cold}\sim 10^4\,$K, $\chi\sim 100$, $\mathcal{M}\lesssim 1$, regime
\begin{equation}
  \label{eq:rccrit}
  r > r_{\rm crit,turb} \equiv 8\,\mathrm{pc} \frac{T_{\rm cl,4}^{5/2}}{P_{3}\Lambda_{\mathrm{mix},-21.4}} {\mathcal{M}} \left( \frac{f(\mathcal{M})}{0.25} \right)^{-1} \left( \frac{\chi}{100} \right)
\end{equation}
where we used $P_{3}=n T / (10^3 \cm^{-3}\,\mathrm{K})$, $\Lambda_{\mathrm{mix},-21.4}\equiv \Lambda(T_{\rm mix}) / (10^{-21.4}\mathrm{erg}\,\cm^3\,\mathrm{s})$, and incorporated the empirical Mach dependence found $f(\mathcal{M})\sim 10^{-0.6 \mathcal{M}}$.
Note that the additional dependency on $\mathcal{M}$ is opposite to what has been found in some classical `cloud crushing' simulations where a higher wind speed implied a \textit{longer} survival time due to compression of the cloud \citep{Scannapieco2015a,Li2019a,Bustard2021}. We attribute this to a more violent fragmentation and subsequent dispersion of the cold gas in our stirring box simulations. For instance, comparing the timescale ratio it takes the cloud to recover after being hit by a shock -- which is of order $t_{\rm grow}$ -- and the time interval between two shocks $\sim t_{\rm eddy}$, $t_{\rm grow}/t_{\rm eddy}\propto \mathcal{M}^{1/2}$, we infer strong disruption at higher Mach numbers. We plan to study this competition between fragmentation / dispersion and cooling driven coagulation in future work.

If Eq.~\eqref{eq:rccrit} is fulfilled, the cold gas can survive and grow, and thus the `initial seed' of cold gas growth needs to be resolved. If a gas cloud, on the other hand, is smaller than $r_{\rm crit,turb}$, it will be destroyed anyway on a short timescale and hence does not need to be resolved numerically in the first place. Below, we compare $r_{\mathrm{crit,turb}}$ to the resolution achieved in a  range of larger scale simulations.

\begin{itemize}
\item The \textit{circumgalactic medium} is multiphase and turbulent (see \S~\ref{sec:CGM_origin}) making our criterion applicable. As has been shown in many (recent) studies, cosmological simulations are unconverged in the total cold gas present in the CGM \citep{Faucher-Giguere2010,VandeVoort2018,Hummels2018,Nelson2020,BennettResolvingshocks2020}.
  As currently, the resolution in the cosmological simulations reaches down to several hundred parsecs, $\gg r_{\rm crit,turb}$ (the values used in Eq.~\eqref{eq:rccrit} are approximately expected for the CGM; e.g., \citealp{Nelson2020}), this is consistent with our expectations. It is noteworthy, though, that \citet{BennettResolvingshocks2020} also find increasing turbulence in their simulated CGM with increasing resolution which might be an important non-converged `loss term' for the cold gas. 
  
\item Sheets and filaments in the \textit{intergalactic medium} can also be multiphase in nature and it has recently been shown that their properties are not converged with current high-resolution cosmological simulations \citep{MandelkerShatteringCosmic2019,Mandelker2021}. Due to the lower pressure in these regions, Eq.~\eqref{eq:rccrit} yields $\sim 80\,$pc (for $\mathcal{M}\sim 1$) which is nearly achievable with current technologies and close to the simulations presented in \citet{MandelkerShatteringCosmic2019,Mandelker2021}.
  
\item It is less clear how this criterion applies for the more violent \textit{interstellar medium} where other dynamic timescales can be shorter than the survival time of the cold gas, and several other phases are present. However, detailed, stratified disk simulations have found that a resolution of $\sim 5\,$pc is required to resolve the warm and hot phases\footnote{For the colder phases a higher resolution is required. \citet{SeifriedSILCCZoomdynamic2017} note that convergence for the mass contained in gas denser than $n>300\cm^{-3}$ is only achieved with a spatial resolution of $0.12\,$pc whereas less dense $n>30\cm^{-3}$ gas only requires $\sim 0.5\,$pc resolution.} \citep{Walch2015,KimThreephaseInterstellar2017} which approximates  what our criterion yields.
  
\item For the \textit{intracluster medium} no systematic resolution study akin to the CGM has been carried out yet. \citet{LiMODELINGACTIVE2014,LiMODELINGACTIVE2014a} simulate isolated clusters and note a convergence in total cold gas mass with a resolution better than a few hundred parsecs which they reach within their refined region along the jet axis. They also note, though, that the number of cold clumps they find is not converged. 
  For a $\dd n/\dd m \propto m^{-2}$  distribution (cf. \S~\ref{sec:mass_dist}), this is expected: one is only logarithmically sensitive to the cutoff mass, so it is possible to converge in total mass even if low mass clumps are unresolved. Similarly, the collapsed halo mass in cosmological simulations converges even if low mass halos (and thus the total number of halos) is unresolved.   
\end{itemize}
Once the ``seed'' of cold gas is resolved, we find that the growth of cold gas is robust to resolution changes.
However, as shown in \S~\ref{sec:results_stochastic}, while resolving $r_{\rm crit}$ is a sufficient criteria for mass convergence, it may not be necessary. What appears to be necessary is sufficient dynamic range between the largest ($r \gg r_{\rm crit}$) cold clump and the cutoff mass, which can be $r_{\rm crit}$ or imposed by numerical resolution. Once again, this is because for a $\dd N/\dd m \propto m^{-2}$, one is only logarithmically sensitive to the cutoff mass. If, on the other hand, seed masses are close the survival boundary $r_{\rm crit}$, then stochasticity and convergence issues become increasingly important. We also caution that our convergence criteria only apply to the production of cold gas via mixing, when `seeds' are already present. Additional resolution requirements may apply to resolve the production of such seeds (via thermal instability, at radiative shocks, etc).

Our conclusions may seem surprising, as one might expect convergence only to occur once the \citet{Field1965} length is resolved, and we do not include thermal conduction in our study, i.e., our multiphase fronts are defined by numerical diffusion. This puzzle has, however, been recently addressed in \citet{Tan2020} using simulations of multiphase mixing layers, and we refer the reader to this study for details (also, see, earlier work \citealp{Ji2018,Fielding2020} focusing on mixing layers for context). The upshot is that convergence in the mass growth is achieved by resolving the outer scale of the mixing process, and not the Field length as the mixing is the `bottleneck' of the mass transfer, and not the thermal (or numerical) diffusion.

\subsection{Implications for the origin and survival of cold gas in the CGM}
\label{sec:CGM_origin}

We know from both classical studies \citep{White1978,birnboimVirialShocksGalactic2003} and cosmological simulations \citep{Shen2013a,Hafen2018,Nelson2020} that the galactic halos of more massive galaxies are filled with a hot $T\sim T_{\rm vir}\gtrsim 10^6\,$K gas. While it is generally hard to impossible to detect this gas directly, we now have even several observational confirmations for this gas phase in nearby galaxies \citep[e.g.,][]{1956ApJ...124...20S,2006ApJ...640..691V}.
Furthermore, the CGM is influenced by cosmological inflows as well as galactic outflows. Both processes seed large scale turbulence within the CGM with $\mathcal{M}\sim 0.1 - 1$ \citep[e.g.,][]{BennettResolvingshocks2020}.

More recent studies have revealed large amounts of cold $\sim 10^4\,$K gas embedded inside this hot halo. Both absorption \citep[see review by][]{Chen2017}  as well as emission studies \citep[e.g,][]{Steidel2010a,Wisotzki2015,Hennawi2015} show that (i) cold gas covers a large region with area covering fractions $f_{\rm A} \gtrsim 0.5$ for $R\lesssim R_{\rm vir}$  \citep[e.g.,][]{Prochaska2017,WildeCGMExtent2021} (ii) however, (at $z\gtrsim 1$) the volume filling factor of this cold gas is very small $\lesssim 10^{-2}$ which means that it resides in dense, cold pockets within the large, much hotter halo; in fact, one can estimate the size of these gas clouds and finds typically $l_{\rm cold}\lesssim 1-100\,$pc (e.g., \citealp{Schayelargepopulation2007,Lau2016}; also see see table 1 of \citealp{McCourt2016} for an overview of the observations and references therein), and (iii) the velocity with respect to the central galaxy of these cold gas pockets is commonly $\sim $a few hundred \kms \citep[][]{Rudie2019a}. 

These detections raise a number of questions regarding the origin, survival and shape (cf. \S~\ref{sec:CGM_cold_gas_shape}) of these cold gas pockets. For the former several suggestions such as in-situ formation via thermal instability,
\citep[e.g.,][]{Sharma2012,Voit2015}, filamentary accretion \citep{Dekel2006,dekelColdStreamsEarly2009,mandelkerInstabilitySupersonicCold2020}, and galactic winds \citep[e.g.,][]{kimSUPERBUBBLESMULTIPHASEISM2016,LiQuantifyingSupernovaedriven2017,Fielding2021} are being discussed in the literature.
The challenge is whether they can deposit sufficient cold gas into the circumgalactic medium to be compatible with the observations.

Using the example of the latter, assuming a cooling wind -- either due to adiabatic expansion \citep{Wang1995,Thompson2015} or via mixing induced cooling (\citealp{Armillotta2017,Gronke2019}; see discussion in \citealp{Farber2021} about using dust \& molecules to differentiate between those cases), one can obtain cold gas material transported into the CGM with mass of order the stellar mass of the galaxy.
It has been argued, however, that the assumed mass loading factors of the wind need to be rather large and / or the observed cold gas velocities are too small \citep{bouchePhysicalPropertiesGalactic2012,Afruni2020}.
Furthermore, in reality galactic winds are not purely radial but also possess some turbulent components that might destroy the cold gas embedded in galactic winds \citep{Vijayan2020,Schneider2020} -- an effect that is usually ignored in `cloud crushing' studies such as ours \citep{Gronke2018,Gronke2019}. 

Furthermore -- independent of its origin -- it is puzzling how this cold gas can survive in such a hostile environment since -- as we saw before -- it should be destroyed on a very short timescale of $\sim 0.8 \mathcal{M} (l_{\rm cold}/ 10\,\mathrm{pc}) (\chi/100)^{1/2} (T / 10^6\,\mathrm{K})^{-1/2}\,$Myr (where we used the length scale and velocities inferred from the observations above).

The results presented in this work could, however, alleviate these problems. Firstly, we showed that cold gas can survive in a turbulent environment. This also implies that, in spite of the presence of turbulence in galactic winds, it is possible to accelerate and entrain cold gas via ram pressure. Superposition of a strong bulk flow with turbulence might create additional effects and further studies are required to investigate this, but we note that the turbulent component in a wind is expected to be within the range probed in this work (see, e.g., figure 21 of \citealp{Schneider2020} suggesting non-radial velocities of $\sim 10-100\kms$, i.e., $\mathcal{M}<1$).

Secondly, as we have shown here, once the cold gas reaches the turbulent CGM, it can not only \textit{survive} but in fact can \textit{grow} exponentially with a mass doubling time of $t_{\rm grow}\sim 530\, (\chi/100) (L / (10 \mathcal{M}\,\mathrm{kpc}))^{1/2} T_{\cl,4}^{3/4} (P_3 \Lambda_{\cl,-23})^{-1/2}\,$Myr where $\Lambda_{\rm cl,-23}\equiv \Lambda(T_{\rm cl})/(10^{-23}\,\mathrm{erg}\,\mathrm{cm}^3\,\mathrm{s}^{-1})$. This means that the initial deposit of cold gas via the processes discussed above can be smaller. For instance, a past starburst would have to transport far less cold material into the CGM where it can then grow and disperse. This is intriguing as high-$z$ galaxies have more commonly wind speeds exceeding their escape velocities \citep[see, e.g., section 5.6 and 6.4][and references therein]{Veilleux2020}.

\subsection{The morphology of cold gas in the CGM: clouds of droplets -- with a core}
\label{sec:CGM_cold_gas_shape}
As discussed in the  previous section, multiple observations indicate the presence of small scale $\lesssim 100\,$pc, cold gas.
\citet{McCourt2016} discussed in a influential paper the potential origin of these droplets via `shattering', a rapid fragmentation process which can explain the formation of such small gas clumps. This paper inspired the picture of a `fog' of cold gas droplets residing in the CGM discussed further in the literature \citep{Liang2018,Sparre2018,Gronke2020}. 
However, the hypothesized size scale of these droplets ($\ll r_{\rm crit}$) renders their survival problematic. Some observations are also hard to explain in this picture. 

Via a novel technique using fast radio bursts (FRBs), \citet{Prochaska2019} claimed to rule out a `mist' of $\sim 0.1\,$pc clouds with volume filling fraction of $10^{-3}$. 
Furthermore, observed quasar absorption line-widths appear inconsistent with a fog spread throughout the (turbulent) galactic halo. Typical turbulent widths of the cold medium are $\sim 10-30\,\kms$ \citep{Crighton2015,Rudie2019a} whereas in a `fog' picture one would to first order expect line widths of order the hot gas turbulent velocity (which for a volume filling fog and transonic turbulence is $\sim v_{\rm circ}$), if the droplets are well entrained (see \S~\ref{sec:res_velocity}). 

Our work addresses these issues. We show that a cold gas cloud can survive in a turbulent environment if $r_{\cl}>r_{\mathrm{crit,turb}}$. 
We also showed that the growth and dynamics of the gas cloud is not monolithic but governed by a complex dynamics of small droplets breaking off. The `cloud' of these droplets has a much higher areal covering fraction than merely the central blob, and because of their proximity they would possess a small relative velocity dispersion (cf. \S~\ref{sec:res_velocity}). Furthermore, these `clouds of droplets' would fill a smaller volume than a `fog' filling the entire halo. It would be interesting to understand how a patchy network of droplets alters existing FRB constraints. 

Thus, the picture of the circumgalactic medium we put forward is maybe more similar to terrestrial clouds: spatially confined droplets which could explain both the low volume filling factor observed as well as the rather small velocity dispersion. The difference is the crucial central `core' of size $\gtrsim r_{\rm crit,turb}$ that ensures long-term survival of the entire system. We plan to investigate these observational implications in future work.

\begin{figure}
  \centering
  \includegraphics[width=\linewidth]{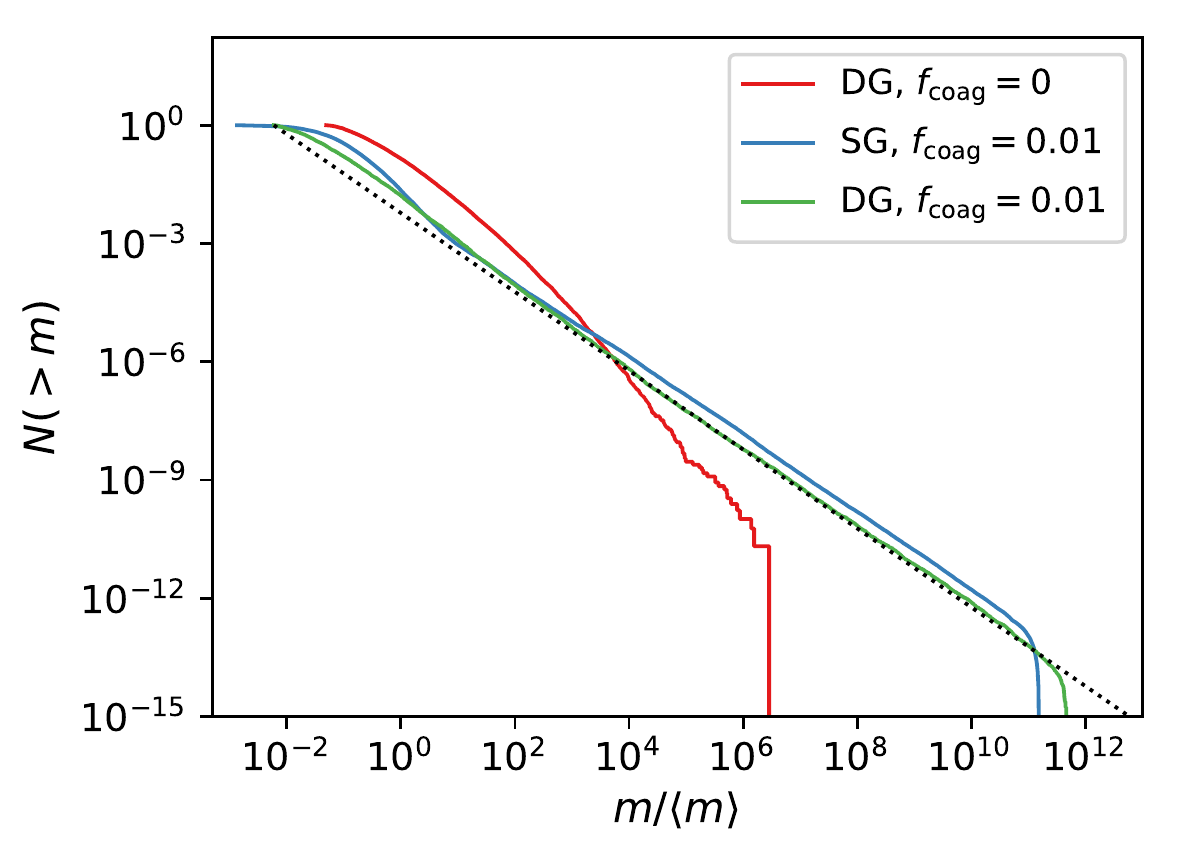}
  \caption{Cumulative droplet mass distributions of our Monte-Carlo simulations with `Double Gaussian' and `Single Gaussian' (DG and SG, respectively) mass redistribution functions as well as different coagulation efficiencies $f_{\rm coag}$ (see \S~\ref{sec:mass_dist} and Appendix~\ref{sec:mc_sims} for details). The dotted line corresponds to $\propto m^{-1}$.}
  \label{fig:MCclumps_dist}
\end{figure}

\subsection{Mass distribution of droplets}
\label{sec:mass_dist}

We found a $\dd N(m)/\dd m\propto m^{-\alpha}$ with $\alpha\sim -2$, that is, corresponding to an approximately constant contribution to mass per logarithmic mass bin (cf. \S~\ref{sec:detail_survival}). This universal power law can be found throughout astrophysics and the natural sciences \citep{NewmanPowerlaws2005}. More relevant to our findings, this droplet mass distribution has been observed in larger scale simulations. For instance, \citet{LiMODELINGACTIVE2014a} found in their simulations of cool-core clusters a log-normal clump mass distribution but note that the turnover at low masses is (similar to ours) due to resolution, and the high mass ($m\in[10^{6.5},\,10^{7.5}]M_{\odot}$) slope follows approximately $\dd N(m)/\dd m\propto m^{-2}$.
Here, we want to discuss why this universal power-law arises in our simulation.

Since power laws in general and in particular the $\dd y/\dd\log x\propto x^{-1}$ power law is so common in nature, a variety of models exist explaining them. In the astrophysical literature, for instance, explaining the $\sim -2$ slope of the stellar initial mass function encompasses a large body of literature \citep[overviews are provided, e.g., in][]{Bonnell2006,Krumholz2014}. Theoretical models invoke a fractal structure of the ISM \citep{Fleck1996}, coagulations of molecular clouds \citep{Silk1979}, as well as fragmentation \citep{Padoan2002,HennebelleAnalyticalTheory2008,Hopkins2012}. 

Here, we adopt a different approach. If we instead focus on droplet growth, which follows approximately $\dot m\propto m$ (cf. \S~\ref{sec:detail_survival}), we can use the mathematical models from the social sciences in which a cumulative power-law distribution with exponent of $-1$ is known as `Zipf's law'. It has been shown that a variety of quantities such as word lengths or city sizes follow it \citep[for a review, see,][]{gabaixPowerLawsEconomics2009}. Below, we sketch a derivation following \citet{Gabaix1999,gabaixPowerLawsEconomics2009}. We plan to study further aspects in future work.

Using normalized droplet masses $\mu_i\equiv m_i/\sum_j m_j$, 
we can write $\dot m \propto m$ as
\begin{equation}
  \mu_i^{(t+1)} = \gamma^{(t+1)}_i \mu^{(t)}_i
  \label{eq:mugrowth}
\end{equation}
where $\gamma_i$ is the growth rate with probability density function $f(\gamma)$. Furthermore, due to the normalized units, we have $\E(\gamma)=\int \gamma f(\gamma)\dd\gamma = 1$. For the cumulative distribution $P(\mu^{(t+1)}>x)$ we can write
\begin{align}
  \label{eq:cumdist_eom}
  P(\mu^{(t)}\gamma^{(t+1)}>x) =& P(\mu^{(t)}>x/\gamma^{(t+1)}) \\
  =& \int\limits_0^{\infty}\dd\gamma\;f(\gamma) P\left( \mu^{(t)}>\frac{x}{\gamma} \right).
\end{align}
Plugging in the power-law ansatz $P(>x)=k/x^\alpha$ for a steady-state distribution one obtains $1=\int\dd \gamma f(\gamma)\gamma^\alpha$ which comparing to $\E(\gamma)=1$ above yields $\alpha=1$.

Importantly, here we assumed such a steady state distribution exists -- which is only the case if Eq.~\eqref{eq:mugrowth} holds for some (high-mass) part of the distribution but not the entire mass range. If it were to hold the over entire mass range, we would end up with a (continuously widening) log-normal distribution \citep[since then Eq.~\eqref{eq:mugrowth} is a standard random walk in log-space][]{gibrat1931inegalits}. That instead the above derived power-law emerges, a dynamically important barrier is necessary, i.e., a lower cutoff mass or a lower mass range for which Eq.~\eqref{eq:mugrowth} does not hold anymore.
Commonly, this barrier is modeled as reflective. Another possibility is an additional additive term to Eq.~\eqref{eq:mugrowth} (which is more important for small masses and becomes negligible for high masses). Both choices imply that the average normalized growth is greater than unity close to the barrier, and hence, smaller than unity in the other parts. This implies that the distribution in the largest parts drifts towards the barrier, in other words, the barrier stays important throughout the evolution.

We recount the heuristic \citet{Gabaix1999,gabaixPowerLawsEconomics2009} argument because it is simple and influential. In fact, as hinted above, it contains a major sleight of hand, since it assumes both a pure power-law and normalized units. Normalized units require a characteristic scale, and by definition a pure power-law is scale free. In realistic scenarios where there is deviation from a pure power-law (in order to have, for instance, a well-defined mean or total mass), the argument above breaks down. In fact, it is possible to show both analytically and via Monte-Carlo simulations that multiplicative processes (plus barriers/additive processes) can produce power-laws with slopes quite different from -1. Factors such as the relative strength of multiplicative and non-multiplicative processes are important in controlling the asymptotic power law slope. We will discuss this in future work. 

In order to be able to study the origin of the clump size distribution in more detail and in particular the nature of the low-mass barrier, we use Monte-Carlo simulations (see Appendix~\ref{sec:mc_sims} for implementation details) in which we simulate the mass trajectories of the Lagrangian tracer particles employed in \S~\ref{sec:detail_survival}. Figure~\ref{fig:MCclumps_dist} shows the resulting mass distribution. We show results using a Gaussian and `Double Gaussian' function for $f(\gamma)$, calibrated to our numerical simulation (we found the exact choice of $f(\gamma)$ to not affect the resulting distribution). Importantly, we also show the impact of coagulation by changing the fraction of transitions to the highest mass $f_{\rm coag}$. From our tracer particle analysis we found $f_{\rm coag}\sim 0.01 - 0.1$ (cf. Fig.~\ref{fig:transitions_vs_size}) which transforms the log-normal distribution (red curve in Fig.~\ref{fig:MCclumps_dist}) closer to the $-1$ power-law. We therefore tentatively conclude that not only the proportional mass growth is crucial in shaping the mass distribution but that coagulation plays an important role. 
Note, however, that Fig. \ref{fig:MCclumps_dist} has a very large dynamic range, which we find necessary for the apparent power-law to develop. This contrasts with the $\sim 3-4$ decades in mass in the simulation data (Fig \ref{fig:mcumhist}). We will investigate this discrepancy in more detail in upcoming work.

\subsection{Caveats}
\label{sec:disc_caveats}

Our study does not address a range of topics which we hope to revisit in future work.
\begin{itemize}
\item \textit{Magnetic fields.} Most astrophysical plasmas are magnetized, and $B$-fields affect the mixing and thus the mass transfer process. Furthermore, they yield non-thermal pressure support which can become large in the cold medium even with initially large plasma $\beta$ due to magnetic compression \citep{Ji2016,Gronke2019}, and they can alter the kinematics due to magnetic draping \citep[e.g.][]{Dursi2008ApJ...677..993D,2015MNRAS.449....2M,Gronke2019,2020ApJ...892...59C}. Adding magnetic fields is a natural extension to this work.
\item \textit{Cosmic rays.} As with magnetic fields, cosmic rays are ubiquitous in astrophysical plasmas and provide a non-thermal pressure support which changes the cooling properties of the gas \citep{Ji2020a,Butsky2020}. Clearly, this is beyond the scope of this work but an interesting future avenue.
\item \textit{Thermal conduction.} Our calculations did not include thermal conduction. Although we expect the mass growth rate not to be  significantly affected by this simplification \citep{Tan2020}, thermal conduction does smooth out cold gas structures smaller than the Field length and, thus, changes the morphology of the multiphase medium \citep[e.g.,][]{Bruggen2016,Armillotta2017} and is worthwhile to study in the future.
\item \textit{Simplified stirring.} While we used numerical techniques commonly used to generate turbulence, how realistic they are is under debate \citep{Federrath2010}. We tested different stirring modes in Appendix~\ref{sec:app_turb_params} and conclude while this has an effect on the mass growth rate it is often smaller than the stochasticity between simulations (cf. \S~\ref{sec:results_stochastic}). A systematic study of the impact of more realistic stirring would therefore require a number of new simulations and is beyond the scope of this work.
\item \textit{Dynamic range.} As in all numerical studies, ours suffers from a finite dynamic range. Particularly worrisome is the small inertial range in turbulent hydrodynamical simulations \citep[e.g.,][]{Dobler2003,Federrath2010}. As shown in Sec.~\ref{sec:results}, we checked the main results of this paper with resolutions ranging from $\sim 4$ to $\sim 64$ cells per cloud radius (cf. \S~\ref{sec:disc_scale}).
\item \textit{Limited domain size.} In order to allow for a reasonable resolution we could only run boxes of size $10-80\,r_\cl$. This is insufficient to study the long-term evolution of the system. We hope to conduct larger simulations in the future to overcome this. 
\end{itemize}
We expect other simplifying assumptions made such as the (initial) cloud geometry affect our results relatively little.

\section{Conclusion}
\label{sec:conclusion}

We investigated the survival and growth process of cold gas in a multiphase turbulent medium. 
Our findings can be summarized as follows:
\begin{enumerate}
\item In a turbulent medium, cold gas survives if it is of size $r \gtrsim r_{\rm crit,turb}\sim 8\,$pc subject to the gas properties (cf. Eq.~\eqref{eq:rccrit}) -- i.e., $t_{\rm cool,mix} \lesssim t_{\rm cc}$ -- and continuously grows.
\item Close to the survival threshold, the long-term evolution is subject to stochastic fluctuations.
\item If the cold gas survives, the mass growth is exponential (due to the continuous fragmentation) and can be described with a simple model (Eq.~\eqref{eq:tgrow_TML_full}).  
\item The clump mass distribution follows the universal $\dd N/\dd m \propto m^{-2}$ power law. We reason
that this is due to the proportional (multiplicative) mass growth as well as the coagulation of smaller clumps with larger clumps, which we test using Monte-Carlo simulations (\S~\ref{sec:mass_dist}).
\item There is good momentum coupling between hot and cold phases and consequently spectroscopy of cool gas give information on hot gas motions (\S~\ref{sec:res_velocity}).
\item These findings lead to a resolution requirement for larger scale simulations of multiphase gases (\S~\ref{sec:disc_scale}) and in particular give insight into the composition of the multiphase galactic halos (\S~\ref{sec:CGM_origin}). The picture supported by our findings is that of a `fog' of droplets surrounding larger clumps ($\gtrsim r_{\rm crit,turb}$) ensuring the survival of the system. This can explain the large area covering but small volume filling fractions and relatively narrow absorption line widths found commonly in the CGM (\S~\ref{sec:CGM_cold_gas_shape}).
\end{enumerate}
More work is required to address the shortcomings of this study  (cf. \S~\ref{sec:disc_caveats}), most notably the long-term evolution and the inclusion of magnetic fields. We will also refine our work on the development of the power law in clump mass distributions.

\section*{Acknowledgments}
We thank the organizers and participants of the KITP ``Fundamentals of Gaseous Halos'' workshop and in particular discussions with Chad Bustard, Hitesh Kishore Das, Ryan Farber, Drummond Fielding, Joe Hennawi, Nir Mandelker, Evan Schneider, and Brent Tan.
This research made use of \texttt{Athena++} \citep{athena,Stone2020}, \texttt{FLASH} \citep{fryxell00}, \texttt{yt} \citep{2011ApJS..192....9T}, \texttt{matplotlib} \citep{Hunter:2007}, \texttt{numpy} \citep{van2011numpy}, and \texttt{scipy} \citep{scipy}.
We acknowledge support from NASA grant NNX17AK58G, 19-ATP19-0205, HST grant HST-AR-15039.003-A, and XSEDE grant TG-AST180036 the Texas Advanced Computing Center (TACC) of the University of Texas at Austin.
MG was supported by by NASA through HST-HF2-51409 awarded by the Space Telescope Science
 Institute, which is operated by the Association of Universities for
 Research in Astronomy, Inc., for NASA, under contract NAS5-26555.
 This research was supported in part by the National Science Foundation under Grant No. NSF PHY-1748958.

\section*{Data Availability}
Data related to this work will be shared on reasonable request to the corresponding author.

 \bibliographystyle{mnras}
 
  \bibliography{refs}

\appendix

\section{Dependence of the mass growth on turbulent parameters}
\label{sec:app_turb_params}

\begin{figure}
  \centering
  \includegraphics[width=\linewidth]{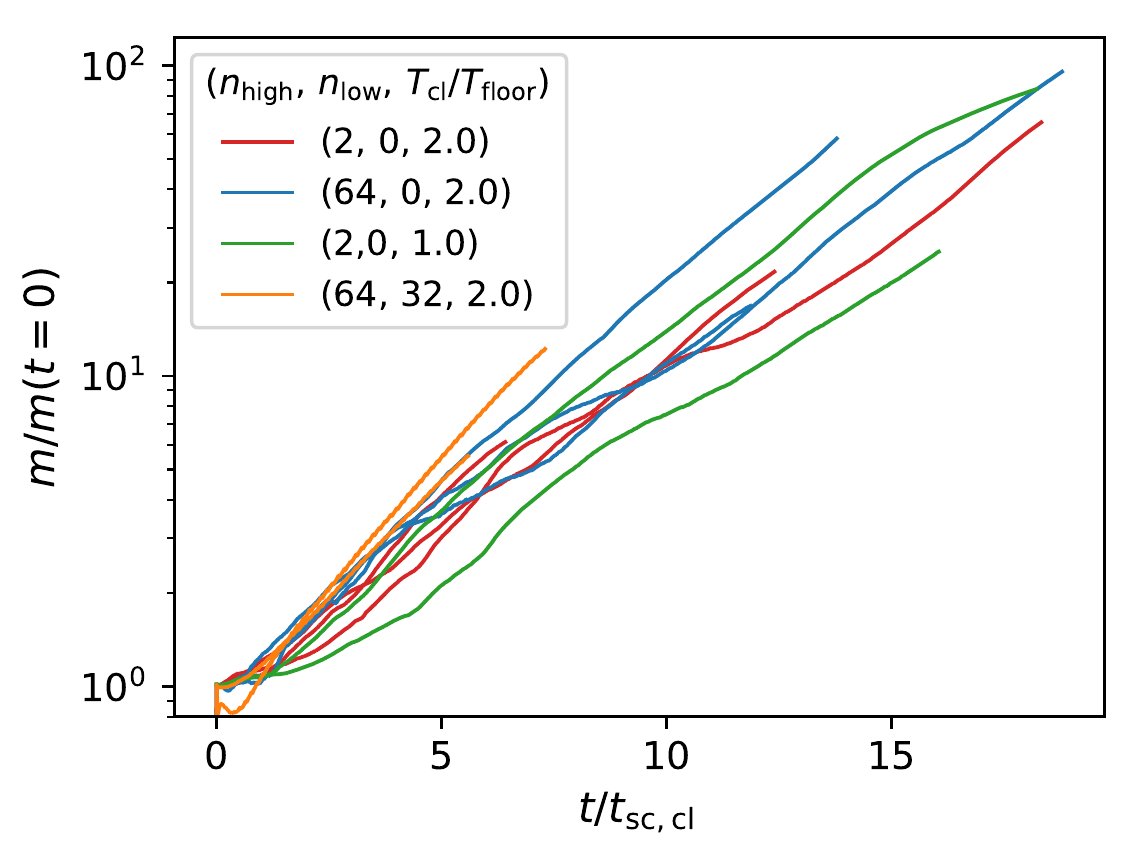}
  \caption{Mass evolution for droplets of size $500\lshatter$ with different driving scales and initial perturbations.}
  \label{fig:m_rd500_nhigh}
\end{figure}

\begin{figure}
  \centering
  \includegraphics[width=\linewidth]{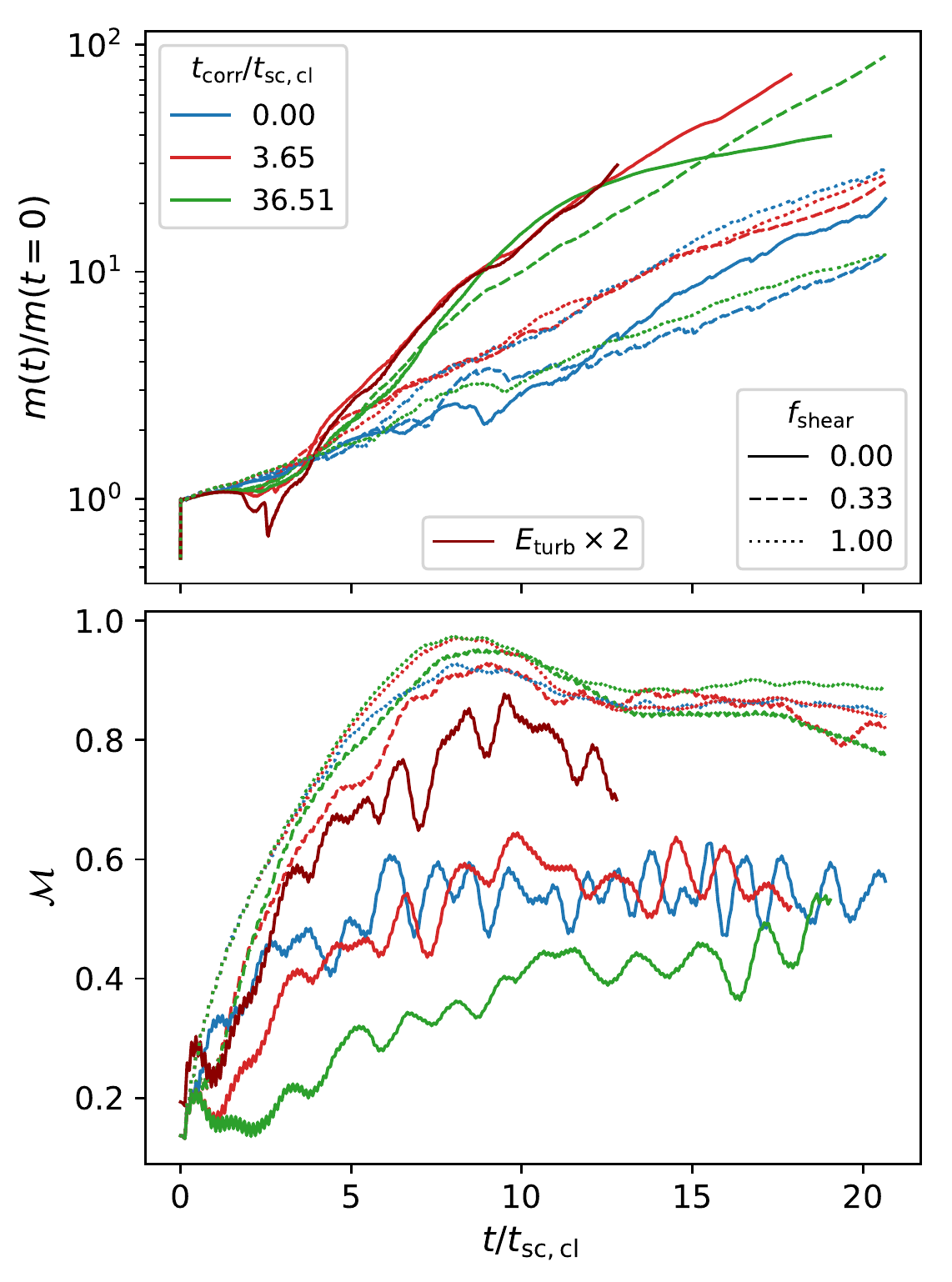}
  \caption{Cold gas mass and Mach number evolution of an initially static box with one cloud of size $r_{\cl}\sim 500\lshatter$ placed inside it for different stirring parameters. The runs were carried out with only $256^3$ elements inside the simulation domain, i.e., $1/8$ of our fiducial resolution. Note that $L/c_{\rm s,hot}\sim 4 t_{\rm sc, cl}$ for reference.}
  \label{fig:fig_stirvar_E1}
\end{figure}

\begin{figure}
  \centering
  \includegraphics[width=\linewidth]{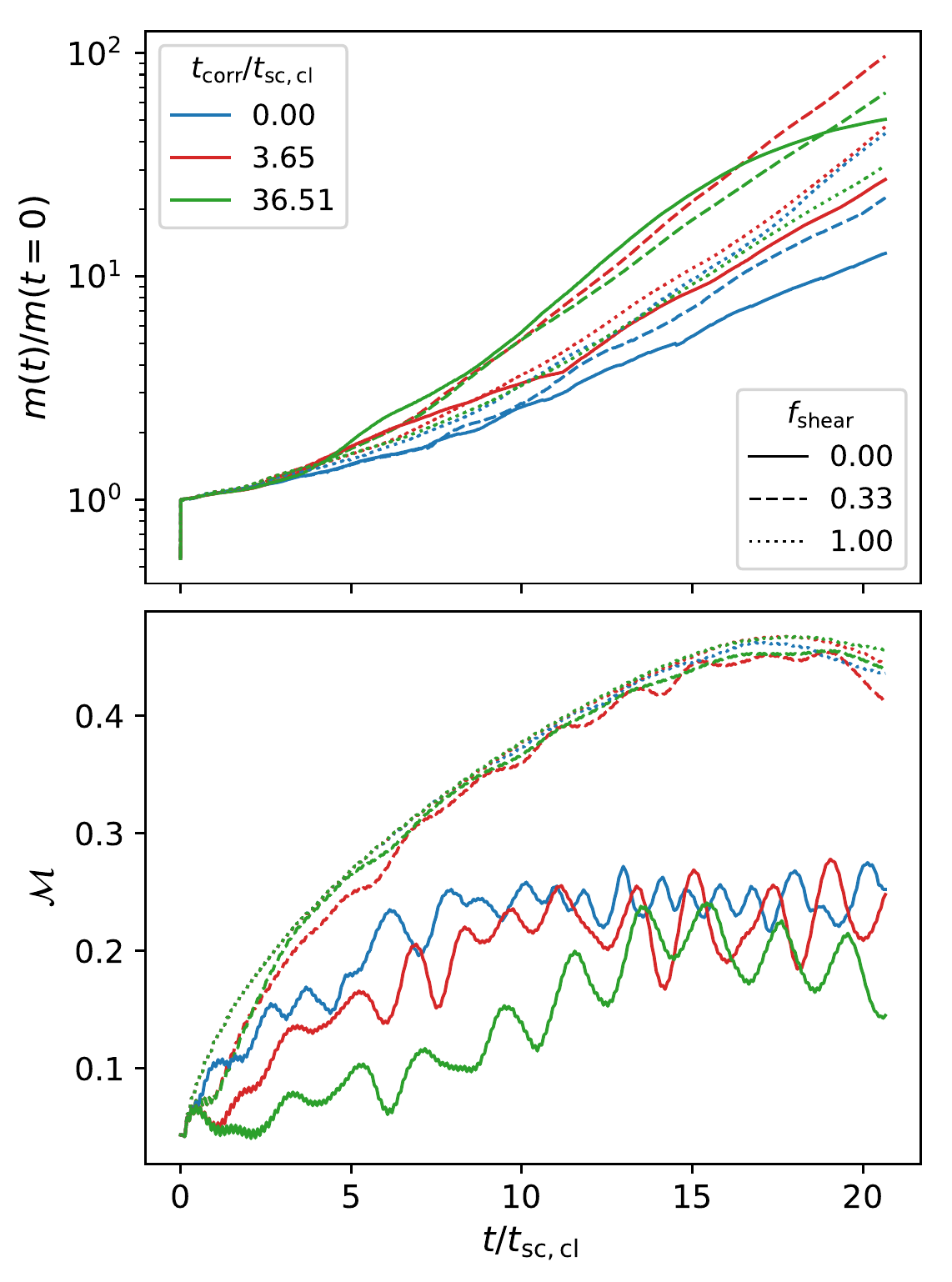}
  \caption{Same as Fig.~\ref{fig:fig_stirvar_E1} but with $1/10$ the stirring energy used.}
  \label{fig:fig_stirvar_E01}
\end{figure}

Figure~\ref{fig:m_rd500_nhigh} shows the impact of the driving scale on the mass growth rate. We changed the driving scale by varying the cutoff wave numbers $n_{\rm low}$ and $n_{\rm high}$ from their default values $0$ and $2$, respectively.
Note that the driving power spectrum slope is chosen to give a $-5/3$ Kolmogorov spectrum. 
The expectation that the mass growth rate is enhanced within the first $t_{\rm eddy}\sim L / v_{\rm turb}$ ($t_{\rm eddy}\sim 7.5 t_{\rm sc,cl}$ in Fig.~\ref{fig:m_rd500_nhigh}) was not fulfilled. This might be due to the fact that the initial mass growth is dominated by the cloud pulsations, since the clouds (initially at $T \sim 2 T_{\rm floor}$) quickly fall out of pressure balance due to rapid cooling. However, in Fig.~\ref{fig:m_rd500_nhigh} we also show runs with $T_{\rm cl} = T_{\rm floor}$, i.e., where the initial pulsations are not present, which show a comparable mass growth rate.

Fig.~\ref{fig:m_rd500_nhigh} also shows two runs in which we only drive on the small scales ($n_{\rm high}=64$, $n_{\rm low}=32$). While this leads to a very different cold gas morphology, the mass growth is consistent which supports our choice of using $l \sim r_{\rm cl}$ to evaluate $t_{\rm grow}$ in Eq.~\eqref{eq:tgrow_TML_full}.

In Fig.~\ref{fig:fig_stirvar_E1} and Fig.~\ref{fig:fig_stirvar_E01} we show the cold gas mass and the Mach number evolution for different turbulent correlation times $t_{\mathrm{corr}}$ and $f_{\mathrm{shear}}$. Before analyzing the results, we want to cautiously remark that these runs are carried out with only $1/8$th the mass resolution compared to our fiducial setup, and also note again the overall stochasticity of the evolution (cf. \S~\ref{sec:results_stochastic}). Firm conclusions would require several higher resolution runs and are beyond the scope of this work. The results presented in Fig.~\ref{fig:fig_stirvar_E1} and Fig.~\ref{fig:fig_stirvar_E01} indicate, however, that:
\begin{enumerate}
\item Purely solenoidal driving ($f_{\mathrm{shear}}=1$) leads to a smaller mass growth. This naively makes sense as compressional driving increases density perturbations and the importance of cooling, as seen in other work (e.g. \citet{Saury2014}). 
\item Uncorrelated driving also leads to slower mass growth, which makes sense since shorter correlation times make it harder for cloudlets to become entrained. 
\end{enumerate}

\section{Monte-Carlo simulations of clump distributions}
\label{sec:mc_sims}

\begin{figure}
  \centering
  \includegraphics[width=\linewidth]{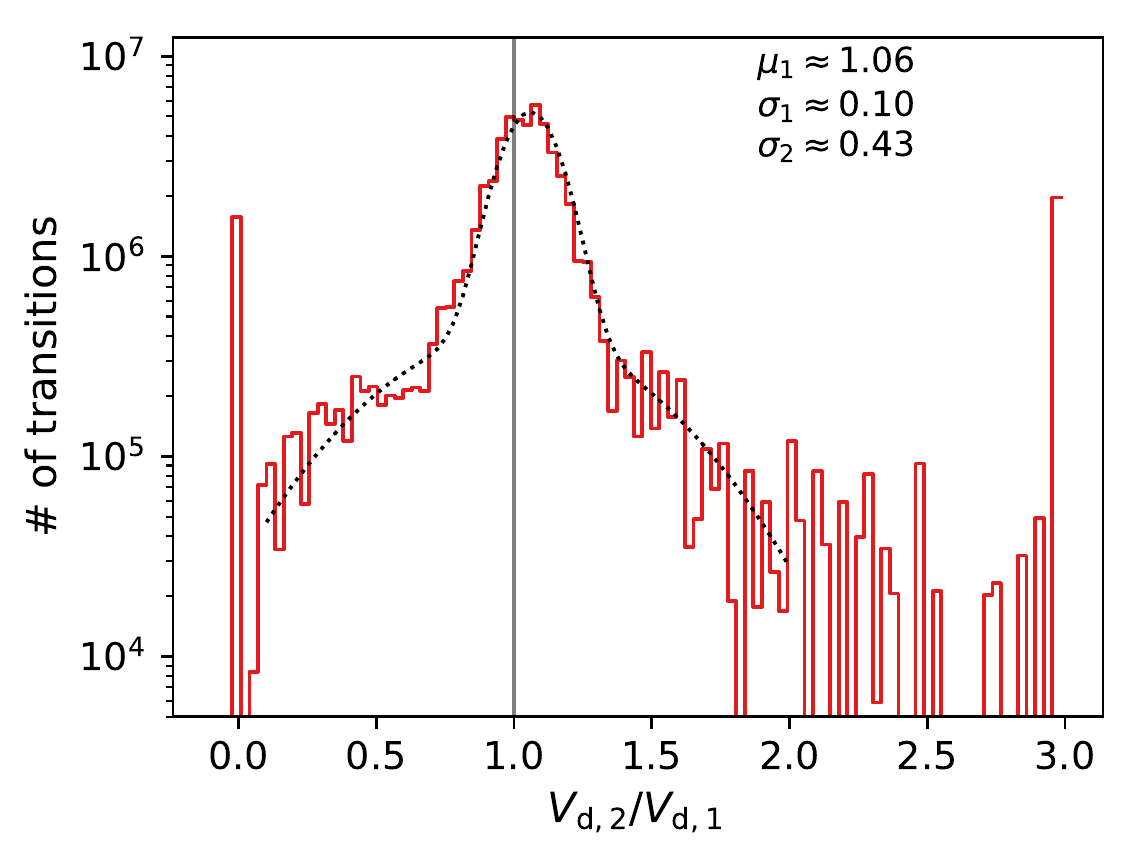}
  \caption{Time integrated probability density function of the clump volume probed by Lagrangian tracer particles. The dotted line  corresponds to our `double Gaussian' model used in the MC simulations. For this figure, we used only particles in clumps of size $V_{\rm d,1}\in (1\pm 0.25)V_{\rm cl}$ to be less prone to boundary effects (but find the distribution to fit universally, cf. \S~\ref{sec:mc_sims}). Apart from that, we did not find a strong variation of $p(V_{\rm d,1}/V_{\rm d,2})$ with $V_{\rm d,1}$. The transitions marked at $V_{\rm d,2}=0$ are into the hot medium.}  
  \label{fig:Vd2Vd1pdf}
\end{figure}

To explore the emergent clump mass distribution, we employ small Monte-Carlo simulations which mimic the mass evolution of the Lagrangian tracer particles in the hydrodynamic simulations. Specifically, we use the following recipe:
\begin{enumerate}
\item Initialize with $N_0$ particles of mass $m_i \sim p(m_0)$ where we use a $\delta$-function or a narrow Gaussian for $p(m_0)$. The results are insensitive to the exact choice of this initial distribution.
\item To model the mass growth from the hot medium which follows $\dot m \propto m$ (see \S~\ref{sec:droplet_dists}), we assign each particle a new mass $f_{m,i} m_i$ with $f_{m,i} \sim p(f_{m,i})$. We calibrate $p(f_{m,i})$ from the hydrodynamical simulation (see below). As we will show below, the mass distribution depends on the choice of $p(f_{m,i})$.
\item To emulate the dynamically important coagulation of clumps, we set $m= \mathrm{max}(m_i)$ for a fraction $f_{\rm coag}$ of particles. This simplistic choice ignores coagulation to intermediate mass particles. However, since in these intermediate mass bins, the mass flux from lower masses is approximately balanced by coagulation. We find that key dynamics are captured by our simple prescription (see details below). We use $f_{\rm coag}\sim 0.01$ as found in the simulations.
\item Remove particles with $m_{i} < m_{\rm cutoff}$ representing clumps lost to the hot medium. 
\item Increase the number of particles by a fixed fraction (the choice of this fraction only affects shot noise) and assign the new particles masses following the current mass distribution. This step simulates the particles changing from hot to cold medium.
\end{enumerate}
The resulting clump mass distribution is then $m^{-1} \dd N(m)/\dd m $, i.e., the particle distribution weighted inversely by the mass.

We tried various functional forms for the growth distribution $p(f_{m,1})$ which we adapted from the simulations. Fig.~\ref{fig:Vd2Vd1pdf} shows the PDF with the `double Gaussian' model consisting of two normal distributions. One can clearly note the dominating ``natural growth'' part for which clumps gain or lose some mass from or to the hot medium, respectively. Furthermore, one can see an extended component consisting of breakup or coagulation (which only make up $\sim 10\%$ of the transitions, also see Fig.~\ref{fig:transitions_vs_size}). We modeled this outer part with another Gaussian, or power-law `wings' fitted to the simulation result but note that the exact shape does not seem to impact the final distribution (cf. \S~\ref{sec:mass_dist} and Fig.~\ref{fig:MCclumps_dist}).

While the extended part of $p(f_{m,1})$ already encapsulates the breakup / coagulations to slightly smaller / larger clumps, the possibility of very small clumps coagulating with much larger ones, and in the process the particles `gaining' a multiple of their original mass, is left out.
This is modeled instead with the $f_{\rm coag}$ parameter introduced above. In fact, as Fig.~\ref{fig:MCclumps_dist} shows including this effect appears necessary to obtaining the $\dd N / \dd m \propto m^{-2}$ power-law observed (see \S~\ref{sec:mass_dist}).

\bsp	
\label{lastpage}
\end{document}